\documentclass[article,amsmath,amssymb,floatfix,aps]{revtex4}
\usepackage{graphicx}
\usepackage{epsfig}
\usepackage{multirow}
\usepackage{color}
\usepackage{cancel}
\usepackage{graphics}

\begin{document}
\title{Simultaneous description of wobbling and chiral properties in even-odd triaxial nuclei}

\author{C. M. Raduta $^{a)}$, A. A. Raduta$^{a), b)}$, R. Poenaru $^{a), c)}$ and Al.H. Raduta $^{a)}$}

\affiliation{$^{a)}$ Department of Theoretical Physics, National Institute of Physics and
  Nuclear Engineering, Bucharest, POBox MG6, Romania}

\affiliation{$^{b)}$Academy of Romanian Scientists, 54 Splaiul Independentei, Bucharest 050094, Romania}

\affiliation{$^{c)}$Doctoral School of Physics, Bucharest University, Soseaua Panduri, nr. 90, Sector 5, 050663, Bucharest, Romania}
\begin{abstract}
A particle-triaxial rigid core Hamiltonian is semi-classically treated. The coupling term  corresponds to a particle  rigidly coupled  to the triaxial core, along a direction that does not belong to any principal plane of the inertia ellipsoid.The equations of motion for the angular momentum components provide a sixth-order algebraic equation for one component and subsequently equations for the other two. Linearizing the equations of motion, a dispersion equation for the wobbling frequency is obtained. The equations of motion are also considered in the reduced space of generalized phase space coordinates. Choosing successively the three axes as quantization axis will lead to analytical solutions for the wobbling frequency, respectively. The same analysis is performed for the chirally transformed Hamiltonian. With an illustrative example one identified wobbling states whose frequencies are mirror image to one another. Changing the total angular momentum I, a pair of twin bands emerged. Note that the present formalism conciliates between the two signatures of triaxial nuclei, i.e., they could coexist for a single nucleus. 
\end{abstract}
\maketitle

\affiliation{$^{b)}$Academy of Romanian Scientists, 54 Splaiul Independentei, Bucharest 050094, Romania}
\setcounter{equation}{0}
\renewcommand{\theequation}{1.\arabic{equation}}
\section{Introduction}
Most of the nuclei from the nuclear chart are axially symmetric and for this reason, the triaxial nuclei were not studied much. The first paper devoted to this issue is that of Davydov and 
Filippov \cite{Davy}.
In refs.\cite{Davy,Terven,WilJean,Marsh,Fra,Bona,Bona2,Bona3} the specific features of triaxial nuclei were studied. Therein two distinct properties are considered as signatures of triaxial nuclei, namely the wobbling motion and chiral doublets \cite{BoMo,Rad016,Petrache96,Petrache06,Frau97,Frau016}. While these features were extensively studied over the recent years
\cite{Ham,Ham1,Tan017,Frau018,Tana018,Rad07,Rad017,Rad018,Rad20,Rad201,Rad2020,Buda},only a separate treatment was considered. A possible justification of this is that up to date the two signatures were experimentally identified  in different nuclei.
However, it is interesting to investigate whether the two properties may be seen in a single nucleus. For this, we have to conciliate two distinct aspects, namely that in the case of chiral states the rotation axis stays outside any principal plane, while in the case of wobbling motion this property is lacking. 

The mentioned signatures of triaxial nuclei are here studied, at a time for even-odd isotopes. A particle-triaxial rotor Hamiltonian is firstly dequantized and then the classical equations of motion for the angular momentum components are explicitly written and an algebraic equation for the stationary angular momentum components are obtained, in Section 2. In Section 3, an equation for the wobbling frequency is derived. The equations of motion in the reduced space of he angular momenta space are considered in Section 4. An illustrative example is treated in Section 5, while the final conclusions are presented in Section 6.
\setcounter{equation}{0}
\renewcommand{\theequation}{2.\arabic{equation}}
\section{Classical description of the  particle triaxial core coupling}
We thus study an odd-mass system  consisting of an even-even core described by a triaxial rotor Hamiltonian $H_{rot}$ and a single j-shell proton moving in a quadrapole deformed mean-field:
\begin{equation}
H_{sp}=\epsilon_j+\frac{V}{j(j+1)}\left[\cos\gamma(3j_3^2-{\bf j}^2)-\sqrt{3}\sin\gamma(j_1^2-j_2^2)\right].
\label{hassp}
\end{equation}
Here $\epsilon_j$ is the single particle energy and $\gamma$, the deviation from the axial symmetric picture.
In terms of the total angular momentum ${\bf I}(={\bf R}+{\bf j}) $ and the angular momentum carried by the odd particle, ${\bf j}$, the rotor Hamiltonian is written as:
\begin{equation}
H_{rot}=\sum_{k=1,2,3}A_k(I_k-j_k)^2.
\label{hrot}
\end{equation}
where $A_k$ are half of the reciprocal moments of inertia associated to the principal axes of the inertia ellipsoid, i.e. $A_k=1/(2{\cal I}_k)$, which are considered as free parameters. 

The expressions for the single particle coupling potential, $H_{sp}$, and the triaxial rotor term, $H_{rot}$, have been previously used by many authors, the first being Davydov \cite{David}.
In the context of rigid coupling of the single particle to the core, the term $H_{sp}$ does not contribute to the equations of motion for the angular momentum components $I_k$, k=1,2,3.

We consider that the maximal moment of inertia (MoI) is ${\cal J}_1$. For a rigid coupling of the odd proton to the triaxial core, we suppose that $\bf{j}$ stays out of any principal plane: 
${\bf j}=(j\cos\theta_0, j\sin \theta_0\cos\varphi_0, j\sin \theta_0\sin\varphi_0)$.
 
We recall that within the liquid drop model (LDM) the odd nucleon may be coupled either to the deformation or to the core angular momentum. Correspondingly, the single particle angular momentum is oriented either to the symmetry axis or to the core's angular momentum \cite{Ring}. These scenarios are reached for weak and strong coupling regimes, respectively. For an intermediate coupling one may meet the situation when ${\bf j}$ lays outside the principal planes. Within a microscopic picture, the orientation of the single particle angular momentum depends on the location of the Fermi level. Thus, when the Fermi level of valence nucleon is located in the lower/upper part of a high-j
subshell, its angular momentum is oriented along the short/long axis of the triaxial core, and in the middle part
with its angular momentum easily aligned with the intermediate axis with the maximum MoI.   When the Fermi level is different from these special cases, the angular momentum of the odd proton might be oriented along a line which is different from the three mentioned axes. In these phenomenological and microscopic contests, it seems reasonable to fix the single particle angular momentum outside any principal plane of the triaxial core.

Note that the linear term in $\hat{I}$ from (\ref{hrot}) looks like the cranking term in the microscopic cranking formalism. According to the pioneering paper of Bengston \cite{Beng} the equations for a general orientation of the cranking term admit a real solution.

In this context we ask ourselves, whether the phenomenological Hamiltonian $(\ref{hrot})$ admits a harmonic solution within a classical treatment.
To this aim we dequantize the model Hamiltonian by replacing the operators $\hat{I}_k$, k=1,2,3 with the classical components of the angular momentum hereafter denoted by $x_k$,k=1,2,3, respectively, and the commutators with the Poisson brackets:
\begin{eqnarray}
&&\hat{I}_k\to x_k,\nonumber\\
&&[,]\to i\{,\} .
\end{eqnarray}
with 'i' denoting the imaginary unity. The Poisson brackets are defined as follows. Let f and g two functions defined on the phase space spanned by the coordinates and conjugate momenta $(q_k,p_k)$
. Then the associated Poisson Bracket is defined as:
\begin{equation}
\{f,g\}=\sum_{k}(\frac{\partial f}{\partial q_k}\frac{\partial g}{\partial p_k}-\frac{\partial f}{\partial p_k}\frac{\partial g}{\partial q_k}).
\end{equation}
According to these rules, the classical energy can be written as:
\begin{eqnarray}
{\cal H}_{rot}&=&AH'+A_1I^2+\sum_{k=1,2,3}A_kj_k^2, \;\;\rm{with}\nonumber\\
H'&=&x_2^2+ux_3^2+2v_1x_1+2v_2x_2+2v_3x_3,\nonumber\\
u&=&\frac{A_3-A_1}{A_2-A_1},\; v_k=-\frac{j_kA_k}{A_2-A_1}, k=1,2,3., A=A_2-A_1.
\label{HandH'}
\end{eqnarray}
Also, the angular momentum components obey:
\begin{equation}
\{x_i,x_j\}=-\epsilon_{i,j,k}x_k.
\end{equation}
where $\epsilon_{i,j,k}$ denotes the three dimensional unity tensor.
Given an operator $O$ defined on the phase space and considered in the interaction representation, this obeys the Heisenberg equation:
\begin{equation}
[O,H]=i\frac{\partial O}{\partial t}.
\end{equation}
According to the dequantization rules the classical counterpart of the above equation is:
\begin{equation}
\{o,H\}=i\frac{\partial o}{\partial t}.
\end{equation}
with $o$ denoting the classical image of $O$.

Since ${\cal H}_{rot}$ and $H'$ differ from each other by one multiplicative and one additive constant, the motion described by ${\cal H}{rot}$ is readily known once that corresponding to $H'$ is given. Due to this feature, it is convenient to deal first with $H'$. Thus,
the equations of motion for the components $x_k$ are:

\begin{eqnarray}
\{x_1,H'\}&=&\stackrel{\bullet}{x}_1=-2\left[x_2x_3(1-u)+v_2x_3-v_3x_2\right],\nonumber\\
\{x_2,H'\}&=&\stackrel{\bullet}{x}_2=-2\left[ux_1x_3-v_1x_3+v_3x_1\right],\nonumber\\
\{x_3,H'\}&=&\stackrel{\bullet}{x}_3=-2\left[-x_1x_2+v_1x_2-v_2x_1\right],\nonumber\\
\{\varphi_1,H'\}&=&\stackrel{\bullet}{\varphi}_1=-2\frac{A_1}{A}(x_1-j_1),\nonumber\\
\{\varphi_2,H'\}&=&\stackrel{\bullet}{\varphi}_2=-2\frac{A_2}{A}(x_2-j_2),\nonumber\\
\{\varphi_3,H'\}&=&\stackrel{\bullet}{\varphi}_3=-2\frac{A_3}{A}(x_3-j_3).
\label{eqmot}
\end{eqnarray}
where the symbol $"\bullet"$ signifies the first derivative with respect to time, while $\varphi_k$ (k=1,2,3) are the conjugate coordinates of $x_k$ (k=1,2,3), respectively.
Note that Eqs.(\ref{eqmot}) are directly obtainable from the Hamilton equations associated to the conjugate coordinates $x_k$ and $\varphi_k$.
The stationary points of the classical trajectories $(x_1(t),x_2(t),x_3(t))|_{t}$ are obtained by cancelling the first derivatives with respect to time of $x_k$, which results in obtaining simple relations between the components $x_k$:
\begin{eqnarray}
&&\frac{v_1}{x_1}-\frac{v_2}{x_2}=1,\nonumber\\
&&\frac{v_1}{x_1}-\frac{v_3}{x_3}=u,\nonumber\\
&&\frac{v_2}{x_2}-\frac{v_3}{x_3}=u-1.
\label{x23ofx1}
\end{eqnarray}
From these relations we can express $x_2$ and $x_3$ in terms of $x_1$ and then insert the result in the angular momentum conservation equation:
\begin{equation}
x_1^2+x_2^2+x_3^2=I^2.
\end{equation}
It results an algebraic equation for the component $x_1$:
\begin{equation}
F(x_1)\equiv\sum_{k=0}^{6}B_kx_1^k=0,
\label{eqx1}
\end{equation}
with the coefficients $B_k$ given by:
\begin{eqnarray}
B_0&=&-I^2v_1^4,\nonumber\\
B_1&=&2v_1^3(1+u)I^2,\nonumber\\
B_2&=&v_1^2\left(v_1^2+v_2^2+v_3^2\right)-v_1^2\left(1+4u+u^2\right)I^2,\nonumber\\
B_3&=&-2\left(v_1^3(1+u)+v_1(v_2^2u+v_3^2)\right)+2v_1u(1+u)I^2,\nonumber\\
B_4&=&v_1^2\left(1+4u+u^2\right)+v_2^2u^2+v_3^2-u^2I^2,\nonumber\\
B_5&=&-2v_1u(1+u),\nonumber\\
B_6&=&u^2.
\end{eqnarray}
We note that, since $B_0\ne 0$, the equation (\ref{eqx1}) does not admit vanishing solutions, which as a matter of fact is a specific feature for the chiral motion. The solution lead to the stationary points 
$(\stackrel{\circ}{x}_1, \stackrel{\circ}{x}_2, \stackrel{\circ}{x}_3)$ for the surface of constant energy:
\begin{equation}
H'=E.
\end{equation}
Among the stationary points some are minima.
Let us denote by $\stackrel{\circ}{x}_1$ a solution of (\ref{eqx1}) which corresponds to the deepest minimum of $H'$. Then from (\ref{x23ofx1}) one gets:
\begin{eqnarray}
\stackrel{\circ}{x}_2&=&\frac{v_2\stackrel{\circ}{x}_1}{v_1-\stackrel{\circ}{x}_1}, \nonumber\\
\stackrel{\circ}{x}_3&=&\frac{v_3\stackrel{\circ}{x}_1}{v_1-u\stackrel{\circ}{x}_1}.
\end{eqnarray}
and thus the minimum point denoted by $\stackrel{\circ}{P}=(\stackrel{\circ}{x}_1,\stackrel{\circ}{x}_2,\stackrel{\circ}{x}_3)$, is fully determined.

\setcounter{equation}{0}
\renewcommand{\theequation}{3.\arabic{equation}}
\section{Small oscillations around the deepest minimum}
The equations of motion for the components $x_k$ are non-linear. However, these can be linearized by replacing one factor of the quadratic terms with the coordinates of the deepest minimum point.
In this way one obtains the following system of linear equations:
\begin{eqnarray}
\{x_1,H'\}=\stackrel{\bullet}{x}_1&=&-\left(2v_2+\stackrel{\circ}{x}_2(1-u)\right)x_3-\left(\stackrel{\circ}{x}_3(1-u)-2v_3\right)x_2,\nonumber\\
\{x_2,H'\}=\stackrel{\bullet}{x}_2&=&-\left(u\stackrel{\circ}{x}_3+2v_3\right)x_1-\left(u\stackrel{\circ}{x}_1-2v_1\right)x_3,\nonumber\\
\{x_3,H'\}=\stackrel{\bullet}{x}_3&=&\left(\stackrel{\circ}{x}2+2v_2\right)x_1-\left(-\stackrel{\circ}{x}_1+2v_1\right)x_2.
\end{eqnarray}
A solution of the linear system of equations may be found by searching for the linear combination:
\begin{equation}
C^*=X_1x_1+X_2x_2+X_3x_3,
\end{equation}
such that the following equation is fulfilled:
\begin{equation}
\{C^*,H'\}=\omega C^*
\end{equation}
This restriction leads to a homogeneous system of linear equations for the coefficients  $X_1,X_2,X_3$. The compatibility condition yields an equation for the frequency $\omega$:
\begin{equation}
\omega^3+3S\omega-2T=0.
\label{eqomeg}
\end{equation} 
with the coefficients given by:
\begin{eqnarray}
3S&=&-\left(2v_1-u\stackrel{\circ}{x}_1\right)\left(\stackrel{\circ}{x}_1-2v_1\right)+\left(u\stackrel{\circ}{x}_3+2v_3\right)\left(2v_3-(1-u)\stackrel{\circ}{x}_3\right)+\left(\stackrel{\circ}{x}_2+2v_2\right)
\left(2v_2+(1-u)\stackrel{\circ}{x}_2\right),\nonumber\\
2T&=&\left(2v_3+u\stackrel{\circ}{x}_3\right)\left(2v_2+(1-u)\stackrel{\circ}{x}_2\right)\left(\stackrel{\circ}{x}_1-2v_1\right)-\left(\stackrel{\circ}{x}_2+2v_2\right)\left((1-u)\stackrel{\circ}{x}_3-2v_3\right)\left(2v_1-u\stackrel{\circ}{x}_1\right).
\end{eqnarray}
The solutions of Eq.(\ref{eqomeg}) are analytically given in Appendix A.
\setcounter{equation}{0}
\renewcommand{\theequation}{4.\arabic{equation}}
\section{The treatment within the reduced space}
Note that the system under consideration exhibits two constants of motion, namely the energy and the angular momentum squared. Furthermore, there is only one independent angular momentum component; adding to this the corresponding conjugate momentum one arrives at a two dimensional phase space, which is conventionally called the {\it the reduced space}. To define this space, it is convenient to use the polar coordinates. We treat separately three cases:
\subsection{Axis 1 is the quantization axis}
In this case the coordinates $x_k$ are:
\begin{equation}
x_1=I\cos\theta_1,\;x_2=I\sin\theta_1\cos\varphi_1,\;x_3=I\sin\theta_1\sin\varphi_1.
\end{equation}
The coordinates $x_2$ and $x_3$ can be expressed in terms of $x_1$ and $\varphi_1$ by replacing first $I\sin\theta_1$ by $\sqrt{I^2-x_1^2}$ and then expanding the square root factor in second order, which results:
\begin{equation}
x_2=\left(I-\frac{1}{2I}x_1^2\right)\cos\varphi_1,\;x_3=\left(I-\frac{1}{2I}x_1^2\right)\sin\varphi_1.
\end{equation}
Inserting these in $H'$ one obtains:
\begin{eqnarray}
H'&=&x_1^2\left(-\cos^2\varphi_1-u\sin^2\varphi_1-\frac{1}{I}(v_2\cos\varphi_1+v_3\sin\varphi_1)\right)\nonumber\\
&+&I^2\cos^2\varphi_1+uI^2\sin^2\varphi_1+2v_2\cos\varphi_1+2v_3\sin\varphi_1+2v_1x_1.
\end{eqnarray}
Note that since the factor $A_2-A_1$ accompanying $H'$ in Eq.(\ref{HandH'}) is positive, the two Hamiltonians, $H'$ and $H_{rot}$, exhibit the same minima. 
Let $(\stackrel{\circ}{x}_1,\stackrel{\circ}{\varphi}_1)$ be the coordinates of the deepest minimum point for $H'$.This minimum point will be determined in next section.
Expanding now $H'$ around the deepest minimum point one obtains:
\begin{eqnarray}
H'&=&\left(-\cos^2\stackrel{\circ}{\varphi}_1-u\sin^2\stackrel{\circ}{\varphi}_1+\frac{v_2}{I}\cos\stackrel{\circ}{\varphi}_1-\frac{v_3}{I}\sin\stackrel{\circ}{\varphi}_1\right)(x_1-\stackrel{\circ}{x}_1)^2\nonumber\\
  &+&\left[\stackrel{\circ}{x}_1^2\left(\cos2\stackrel{\circ}{\varphi}_1-u\cos2\stackrel{\circ}{\varphi}_1+\frac{v_2}{2I}\cos\stackrel{\circ}{\varphi}_1+\frac{v_3}{2I}
\sin\stackrel{\circ}{\varphi}_1\right)\right.\nonumber\\
  &-&\left.(1-u)I^2\cos2\stackrel{\circ}{\varphi}_1-v_2I\cos\stackrel{\circ}{\varphi}_1-v_3I\sin\stackrel{\circ}{\varphi}_1\right](\varphi_1-\stackrel{\circ}{\varphi}_1)^2.
\end{eqnarray}

It is worth noting that the above Taylor expansion is lacking the first order as well as the mixed second order derivatives. The reason is that the first order derivatives are vanishing in a minimum point. As for the other mising term this is omitted since violates the time reversal symmetry, being linear in momenta

For positive coefficients of the deviations squared, the above equation describes a harmonic oscillator of frequency:
\begin{eqnarray}
\omega^{(1)}&=&2\left[\left(\cos^2\stackrel{\circ}{\varphi}_1+u\sin^2\stackrel{\circ}{\varphi}_1+\frac{1}{I}(v_2\cos\stackrel{\circ}{\varphi}_1+v_3\sin\stackrel{\circ}{\varphi}_1)\right)\right.
\nonumber\\&\times&\left.\left((I^2-\stackrel{\circ}{x}_1^2)(1-u)
\cos2\stackrel{\circ}{\varphi}_1+\frac{1}{I}(I^2-\frac{\stackrel{\circ}{x}_1^2}{2})(v_2\cos\stackrel{\circ}{\varphi}_1+v_3\sin\stackrel{\circ}{\varphi}_1\right)\right]^{1/2}.
\label{omeg1}
\end{eqnarray}
\subsection{Axis 2 is the quantization axis}
In this case we choose the coordinates as:
\begin{equation}
x_2=I\cos\theta_2,‌\;x_3=I\sin\theta_2\cos\varphi_2,\;x_1=I\sin\theta_2\sin\varphi_2.
\end{equation}
Following the procedure from the previous subsection, the energy function $H'$ is expressed in terms of the conjugate coordinate $(x_2,\varphi_2)$:
\begin{eqnarray}
H'&=&x_2^2+2v_2x_2-ux_2^2\cos^2\varphi_2-\frac{1}{I}x_2^2(v_1\sin\varphi_2+v_3{I}\cos\varphi_2)\nonumber\\
&+&uI^2\cos^2\varphi_2+2v_1\sin\varphi_2+2v_3I\cos\varphi_2.
\end{eqnarray}
The deepest minimum is reached in $( \stackrel{\circ}{x}_2,\stackrel{\circ}{\varphi}_2)$, to be determined in next section.

Expanding $H'$ around this minimum one gets:
\begin{eqnarray}
H'&=&\left[1-u\cos^2\stackrel{\circ}{\varphi}_2-\frac{1}{I}\left(v_1\sin\stackrel{\circ}{\varphi}_2+v_3\cos\stackrel{\circ}{\varphi}_2\right)\right](x_2-\stackrel{\circ}{x}_2)^2\nonumber\\
&+&
\left[(\stackrel{\circ}{x}_2^2-I^2)u\cos2\stackrel{\circ}{\varphi}_2+(\frac{\stackrel{\circ}{x}_2^2}{2I}-I)\left(v_1\sin\stackrel{\circ}{\varphi}_2+v_3\cos\stackrel{\circ}{\varphi}_2\right)
\right](\varphi_2-\stackrel{\circ}{\varphi}_2)^2.
\end{eqnarray}
If the coefficients of the variation of $(x_2-\stackrel{\circ}{x}_2)^2$ and $(\varphi_2-\stackrel{\circ}{\varphi}_2)^2$, respectively, are positive,the Hamiltonian $H'$ describes a linear oscillator of frequency:
\begin{eqnarray}
\omega^{(2)}&=&2\left[\left(1-u\cos^2\stackrel{\circ}{\varphi}_2-\frac{1}{I}\left(v_1\sin\stackrel{\circ}{\varphi}_2+v_3\cos\stackrel{\circ}{\varphi}_2\right)\right)\right.\nonumber\\
&\times&\left.\left((\stackrel{\circ}{x}_2^2-I^2)u\cos2\stackrel{\circ}{\varphi}_2+(\frac{\stackrel{\circ}{x}_2^2}{2I}-I)\left(v_1\sin\stackrel{\circ}{\varphi}_2+v_3\cos\stackrel{\circ}{\varphi}_2
\right)\right)\right]^{/2}.
\label{omeg2}
\end{eqnarray}
\subsection{Axis 3 is the quantization axis}
To this case the following polar coordinates correspond:
\begin{equation}
x_3=I\cos\theta_3,\;x_1=I\sin\theta_3\cos\varphi_3,\;x_1=I\sin\theta_3\sin\varphi_3.
\end{equation}
Inserting these coordinates in the  expression of $H'$, it results:
\begin{eqnarray}
H'&=&\left(I^2-x_3^2\right)\sin^2\varphi_3+ux_3^2+v_3x_3\nonumber\\
&+&2I(v_1\cos\varphi_3+v_2\sin\varphi_3)-\frac{x_3^2}{I}(v_1\cos\varphi_3+v_2\sin\varphi_3).
\end{eqnarray}
The  minimum of $H'$ is given by $(\stackrel{\circ}{x}_3,\stackrel{\circ}{\varphi}_3)$, which is introduced in next section.

The quadratic expansion of $H'$ around the deepest minimum point $(\stackrel{\circ}{x}_3,\stackrel{\circ}{\varphi_3})$ is:
\begin{eqnarray}
H'&=&\left(-\sin^2\stackrel{\circ}{\varphi}_3+u-\frac{1}{I}(v_1\cos\stackrel{\circ}{\varphi}_3+v_2\sin\stackrel{\circ}{\varphi}_3))\right)(x_3-\stackrel{\circ}{x}_3)^2\nonumber\\
&+&\left((I^2-\stackrel{\circ}{x}_3^2\cos2\stackrel{\circ}{\varphi}_3+(\frac{\stackrel{\circ}{x}_3^2}{2I}-I)(v_1\cos\stackrel{\circ}{\varphi}_3+v_2\sin\stackrel{\circ}{\varphi}_3)\right)
(\varphi_3-\stackrel{\circ}{\varphi}_3)^2
\end{eqnarray}
Under circumstances that the second derivatives of $H'$ with respect to $x_3$ and $\varphi_3$, respectively, are positive, $H'$ describes a linear oscillator having the frequency:
\begin{eqnarray}
\omega^{(3)}&=&2\left[\left(-\sin^2\stackrel{\circ}{\varphi}_3+u-\frac{1}{I}(v_1\cos\stackrel{\circ}{\varphi}_3+v_2\sin\stackrel{\circ}{\varphi}_3))\right)\right.\nonumber\\
&\times&\left.\left((I^2-\stackrel{\circ}{x}_3^2)\cos2\stackrel{\circ}{\varphi}_3+(\frac{\stackrel{\circ}{x}_3^2}{2I}-I)(v_1\cos\stackrel{\circ}{\varphi}_3+v_2\sin\stackrel{\circ}{\varphi}_3)\right)\right]^{1/2}.
\label{omeg3}
\end{eqnarray}
If the frequencies (\ref{omeg1}), (\ref{omeg2}), (\ref{omeg3}) are all real, then these describe the wobbling frequencies for the motion along the axes 1,2,3, respectively.
It is interesting to see what is the relations between the   frequencies given in subsections A, B and C and the solutions of the cubic equation (\ref{eqomeg}) from the previous section. This issue will be pointed out in next section.

In the space of angular momentum, a chiral transformation is equivalent to the space inversion operation, i.e. $C={\bf I}\to -{\bf I}$. Due to the linear terms in angular momentum components, the Hamiltonian $H'$ is not invariant to chiral transformations. On the other hand, if there exists an operator $O$ which anti-commutes with a given Hamiltonian $H$,
\begin{equation}
\{H,O\}=0,
\label{Hanticom}
\end{equation}
then, if $\Psi$ is an eigenfunction of $H$ corresponding to the eigenvalue $\lambda$, it results that $O\Psi$ is also an eigenfunction of $H$ with the eigenvalue $-\lambda$.
Therefore, the eigenvalues $\lambda$ and -$\lambda$ are mirror images of one another.
In our case $H'$ is a sum of two terms, one invariant, $H_1$, and another non-invariant, $H_2$, to chiral transformations C. The non-invariant term $H_2$ and the transformation C satisfy 
Eq.(\ref{Hanticom}). Due to this feature the eigenvalues of $H'$ are mirror images of those for $CH'C^{-1}$.The two sets of energies define the so called chiral bands.
We note that $CH'C^{-1}$ is obtained from $H'$ by changing $v_k\to  -v_k$, which results that the wobbling frequencies, $\omega^{(k)}_{ch}$, built up with $CH'C^{-1}$ are obtained from those obtained with $H'$ with  the transformation $v_k\to  -v_k$.Thus, we have:
\begin{eqnarray}
\omega^{(1)}_{ch}&=&2\left[\left(\cos^2\stackrel{\circ}{\varphi}_1+u\sin^2\stackrel{\circ}{\varphi}_1-\frac{1}{I}(v_2\cos\stackrel{\circ}{\varphi}_1+v_3\sin\stackrel{\circ}{\varphi}_1)\right)\right.\nonumber\\
&&\left.\times\left((I^2-\stackrel{\circ}{x}_1^2)(1-u)
\cos2\stackrel{\circ}{\varphi}_1-\frac{1}{I}(I^2-\frac{\stackrel{\circ}{x}_1^2}{2})(v_2\cos\stackrel{\circ}{\varphi}_1+v_3\sin\stackrel{\circ}{\varphi}_1\right)\right]^{1/2},\nonumber\\
\omega^{(2)}_{ch}&=&2\left[\left(1-u\cos^2\stackrel{\circ}{\varphi}_2+\frac{1}{I}\left(v_1\sin\stackrel{\circ}{\varphi}_2+v_3\cos\stackrel{\circ}{\varphi}_2\right)\right)\right.\nonumber\\
&&\times\left.\left((\stackrel{\circ}{x}_2^2-I^2)u\cos2\stackrel{\circ}{\varphi}_2-(\frac{\stackrel{\circ}{x}_2^2}{2I}-I)\left(v_1\sin\stackrel{\circ}{\varphi}_2+v_3\cos\stackrel{\circ}{\varphi}_2\right)
\right)\right]^{/2},\nonumber\\
\omega^{(3)}_{ch}&=&2\left[\left(-\sin^2\stackrel{\circ}{\varphi}_3+u+\frac{1}{I}(v_1\cos\stackrel{\circ}{\varphi}_3+v_2\sin\stackrel{\circ}{\varphi}_3))\right)\right.\nonumber\\
&&\times\left.\left((I^2-\stackrel{\circ}{x}_3^2\cos2\stackrel{\circ}{\varphi}_3-(\frac{\stackrel{\circ}{x}_3^2}{2I}-I)(v_1\cos\stackrel{\circ}{\varphi}_3+v_2\sin\stackrel{\circ}{\varphi}_3)\right)
\right]^{1/2}.
\label{omegch}
\end{eqnarray}
The spectrum of $H_{rot}$ is:
\begin{eqnarray}
E^{(1)}_{I,n}&=&(A_2-A_1)\left(H'^{(1)}_{min}+\omega^{(1)}(n+\frac{1}{2})\right)+\sum_{k=1,2,3}A_kj_k^2,\nonumber\\
E^{(2)}_{I,n}&=&(A_2-A_1)\left(H'^{(2)}_{min}+\omega^{(2)}(n+\frac{1}{2})\right)+\sum_{k=1,2,3}A_kj_k^2,\nonumber\\
E^{(3)}_{I,n}&=&(A_2-A_1)\left(H'^{(3)}_{min}+\omega^{(3)}(n+\frac{1}{2})\right)+\sum_{k=1,2,3}A_kj_k^2, n=0,1,2,....
\end{eqnarray}
The mirror images of these energies through the Chiral transformation are:
\begin{eqnarray}
E^{(1)}_{ch,I,n}&=&(A_2-A_1)\left(H'^{(1)}_{ch,min}+\omega^{(1)}_{ch}(n+\frac{1}{2})\right)+\sum_{k=1,2,3}A_kj_k^2,\nonumber\\
E^{(2)}_{ch,I,n}&=&(A_2-A_1)\left(H'^{(2)}_{ch,min}+\omega^{(2)}_{ch}(n+\frac{1}{2})\right)+\sum_{k=1,2,3}A_kj_k^2,\nonumber\\
E^{(3)}_{ch,I,n}&=&(A_2-A_1)\left(H'^{(3)}_{ch,min}+\omega^{(3)}_{ch}(n+\frac{1}{2})\right)+\sum_{k=1,2,3}A_kj_k^2, n=0,1,2,....
\end{eqnarray}
The notations $H'^{(k)}_{min}$ and $H'^{(k)}_{ch,min}$ are used for minimal energy when the axis "k" is the quantization axis. According to Eq.(\ref{omeg1}), (\ref{omeg2}), (\ref{omeg3}),
and (\ref{omegch}), the energies are angular momentum dependent. For a given $n$ and I=$\alpha$+2n with $\alpha$  being the signature, the set of energies $E^{(k)}_{I,n}$ defines a wobbling band, while
$E^{(k)}_{ch,I,n}$ the chiral partner band. In this way we found out a set of states which are simultaneously of wobbling and chiral character.
The wobbling motion is conciliated with the ingredient that the rotation axis is outside the principal planes.

Before closing this section few details about the chiral transformations are necessary. The chiral transformation is bringing a right handed frame to a left handed frame. For example, if $(x_1, x_2, x_3)$ is right handed, then $(-x_1,- x_2, -x_3)$ is left handed. The components of the angular momentum ${\bf j}$ in the two frames are $(j_1,j_2,j_3)$ and
$(-j_1,-j_2,-j_3)$, respectively. The correspondence between ${\bf j}$ and ${\bf -j}$ is a chiral transformation (C) in the space of angular momenta. This definition allows us to study the Hamiltonians whose eigenvalues are sensitive to the change of the rotation sense for the system under consideration. Here, this type of chiral transformation is studied. We note that the transformation
$C_3=(j_1,j_2,j_3)\to (j_1,j_2,-j_3)$ is also chiral. So are $C_1=(j_1,j_2,j_3)\to (-j_1,j_2,j_3)$ and $C_2=(j_1,j_2,j_3)\to (j_1,-j_2,j_3)$. Moreover, the wobbling frequency for the chiral image of $H_{rot}$ through the transformations $C_k$ is obtained from Eqs.(4.5), (4.9) and (4.13) by changing $v_k$ to $-v_k$, respectively. Obviously,the transformations $C_k$, k=1,2,3 are related to the chiral transformation C by:
\begin{equation}
C=C_1C_2C_3.
\end{equation}

\setcounter{equation}{0}
\renewcommand{\theequation}{5.\arabic{equation}}
\section{An illustrative example}

Here we consider an odd particle from the single particle orbit $j=i_{13/2}\hbar$ moving around a triaxial rigid rotor core with the moments of inertia (MoI):
\begin{equation}
({\cal J}_1, {\cal J}_2, {\cal J}_3)= (60, 20, 30)\hbar^2MeV^{-1}.
\end{equation}
The composite system moves in a state of angular momentum $I=35/2\hbar$. The odd particle is rigidly coupled to the core such that its angular momentum orientation is outside the principal planes of the triaxial ellipsoid. Thus, the polar coordinates of the vector ${\bf j}$ are: ${\bf j}=(j,\theta_0,\varphi_0)=(13/2,\pi/4,\pi/4)$. 
The stationary points for the equations of motion for the classical angular momentum components $x_k$, k=1,2,3, obey a set of equations which leads to an algebraic sixth-order equation for $x_1$
,i.e $F(x_1,I)=0$. The function $F(x_1,I)$ is pictorially given in Fig.1.
\begin{figure}[ht!]
\includegraphics[width=0.4\textwidth]{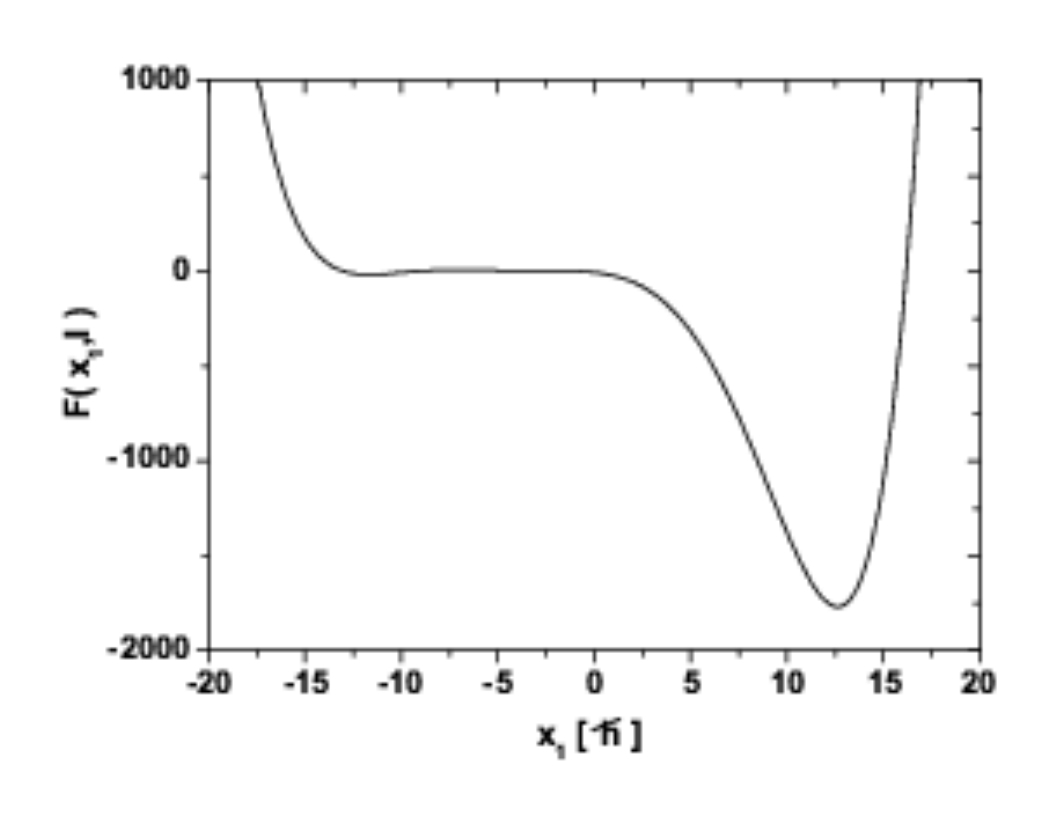}\includegraphics[width=0.35\textwidth]{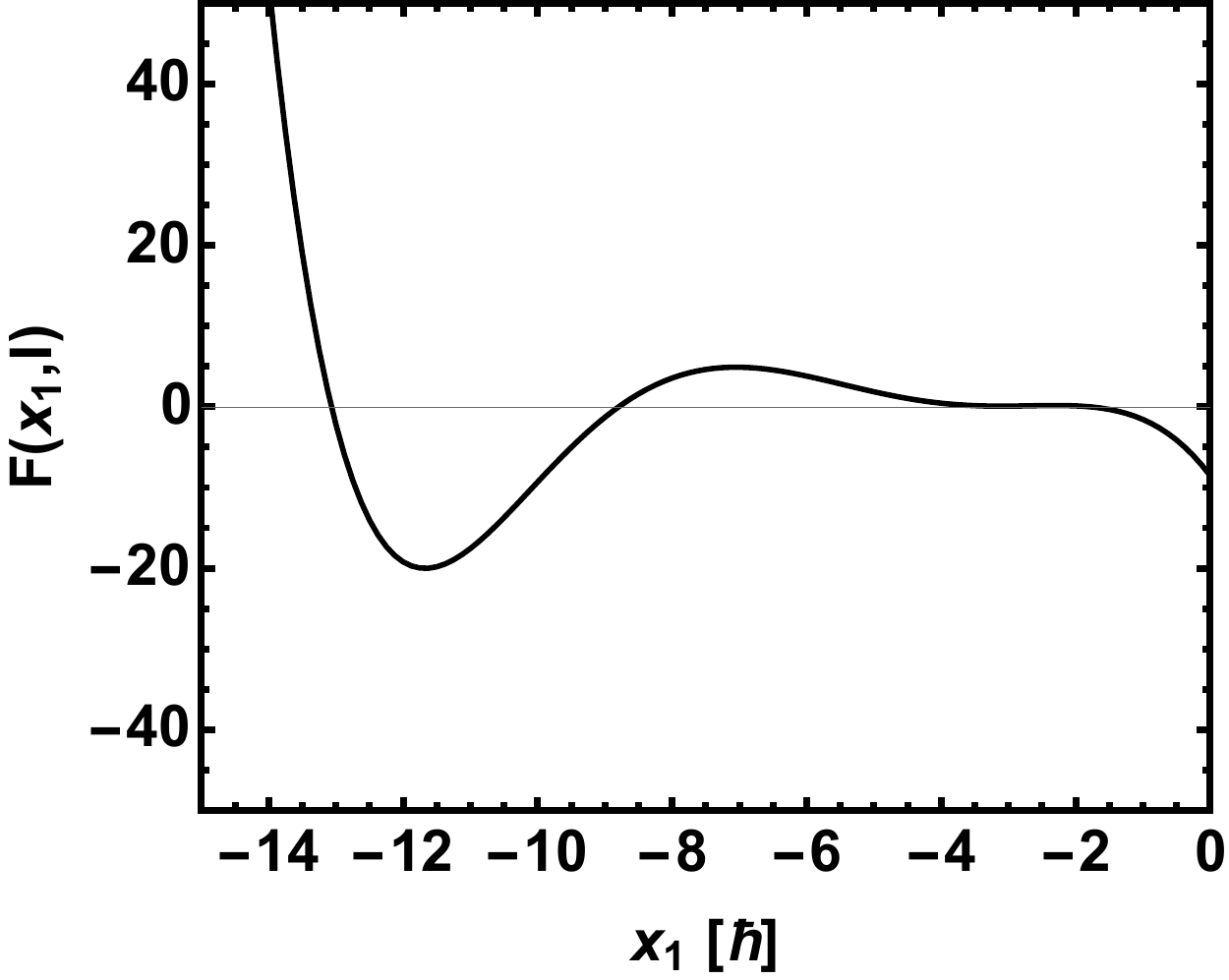}
\caption{The function $F(x_1,I)$  given by Eq.(2.9), is represented (left panel) as function of $x_1$ for the angular momentum $I=35/2\hbar$. The same function is plotted within a shorter interval (right panel) such that the first three solutions are visible.}
\label{Fig.1}
\end{figure}
This function admits four real solutions for $x_1$:$-13.062; -8.811; -1.81; 16.185 [\hbar]$. Making use of relations expressing $x_2$ and $x_3$ in terms of $x_1$ one arrives at the final result:
\begin{eqnarray}
(\stackrel{\circ}{x}_1, \stackrel{\circ}{x}_2, \stackrel{\circ}{x}_3)=\left(\begin{matrix} & -13.062&   5.916&  10.029\\
                                   & -8.811 &  6.595&  13.588\\
                                   & -1.810& 16.886 &  -4.223\\
                                    & 16.185&   4.269&   5.062\end{matrix}  \right)\hbar.
\label{stavect}
\end{eqnarray}

To the four stationary vectors  the following  classical energies correspond:
\begin{eqnarray}
H_{rot}=\left(\begin{matrix}&axis-1&axis-2&axis-3\\
                            &3.542&3.027&2.887 \\
                            &3.559&3.093&2.839\\
                            &5.921&4.920&5.793\\
                            &1.200&1.452&1.424\end{matrix}\right) \rm{MeV}.
\label{energ}
\end{eqnarray}
For example, for the stationary angular momenta components from the row 1 of Eq. (\ref{stavect}), the energies of the row 1 from Eq.(\ref{energ}) correspond, for the situations when the quantization axes are the axis-1,axis-2 and axis-3, respectively.

The frequencies characterizing the linearized equations of motion satisfy a third order algebraic equation. The minimum value for the energy $H_{rot}$ when ${\bf I}\parallel {\bf j}$ ,
i.e. when the two angular momenta are aligned, is 1.765 MeV. 
 With this data there exists a real solution for the wobbling frequency:
\begin{equation}
\omega = 0.362 {\rm MeV.}
\end{equation}
The chiral partner state has the frequency equal to 3.651 MeV.

Furthermore, we studied the equations of motion for $H'$ in the reduced space of generalized phase space coordinates.
\begin{figure}[h!]
\begin{center}
\includegraphics[width=0.5\textwidth]{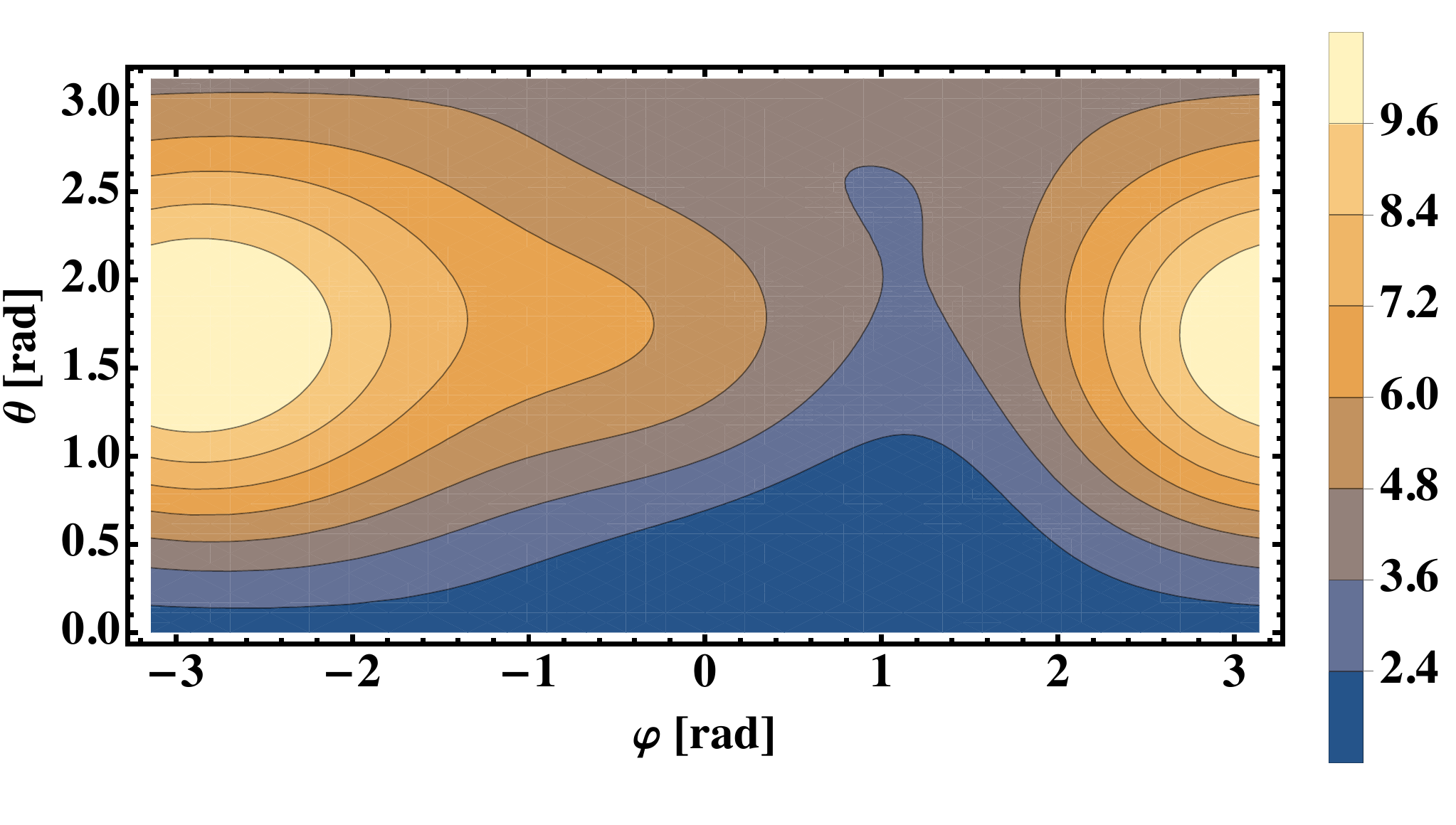}
\end{center}
\caption{(Color online)  The energy function is $H_{rot}$ given by Eq. (2.2)
 as a function of the polar coordinates. Contour plot when the axis-1 is the quantization axis.}
\label{Fig.2}
\end{figure}

\begin{figure}[h!]
\begin{center}
\includegraphics[width=0.5\textwidth]{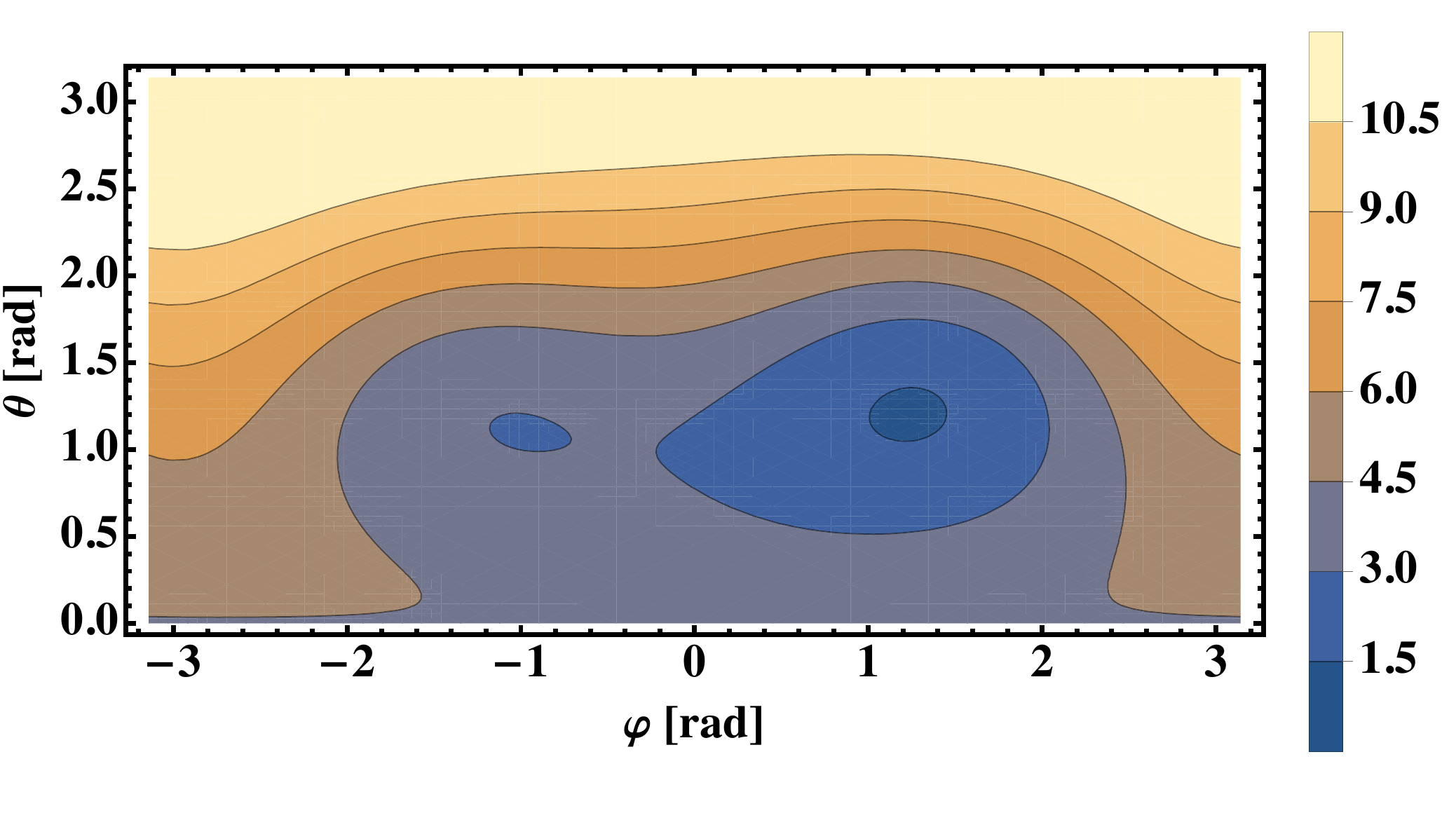}
\end{center}
\caption{(Color online) The energy function is $H_{rot}$ given by Eq. (2.2) as a function of the polar coordinates. Contour plot with the axis-2 as the quantization axis.}
\label{Fig.3}
\end{figure}

\begin{figure}[ht!]
\begin{center}
\includegraphics[width=0.5\textwidth]{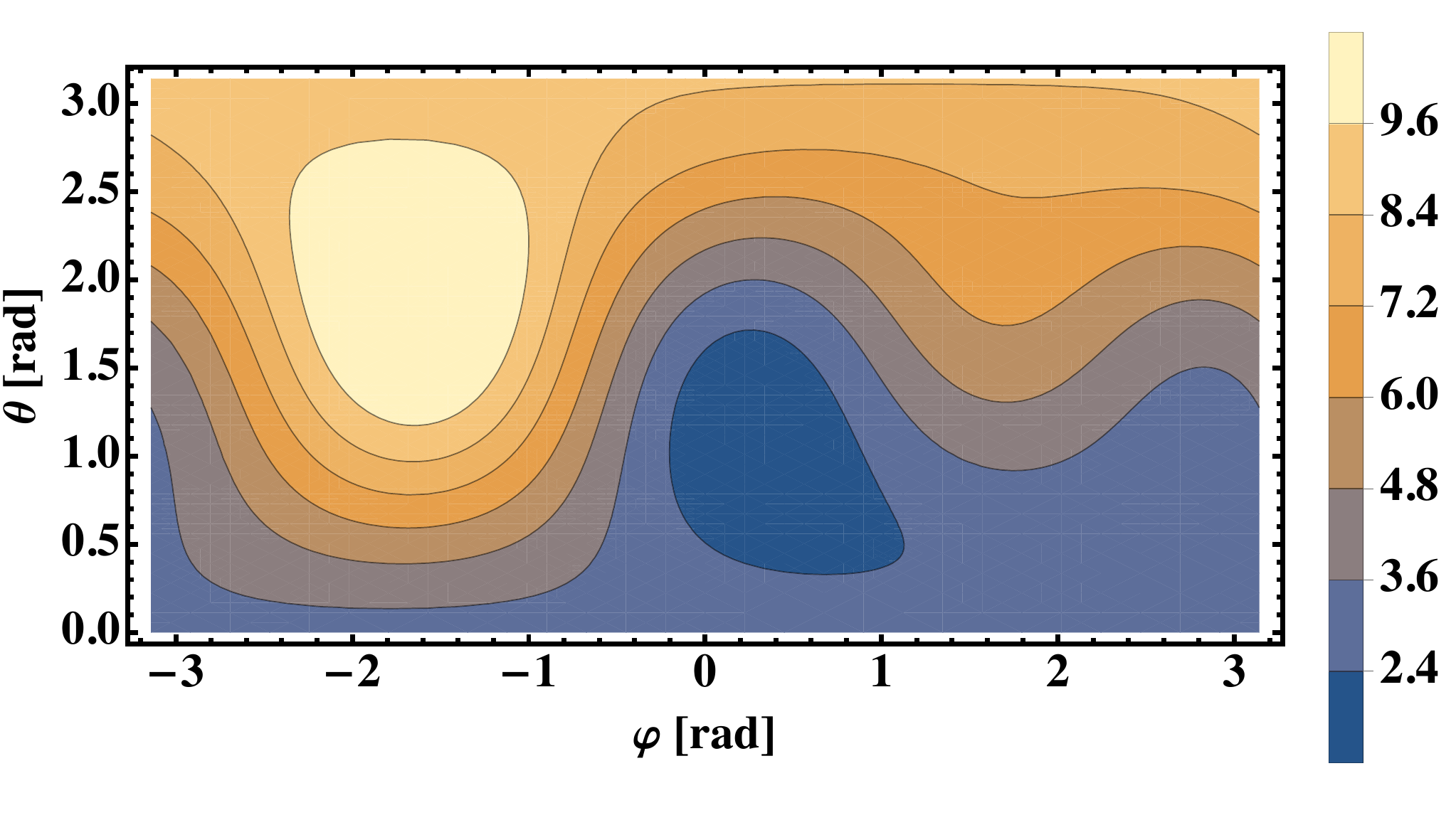}
\end{center}
\caption{(Color online) The energy function is $H_{rot}$ given by Eq. (2.2) as a function of the polar coordinates. Contour plot with the axis-3 as the quantization axis.}
\label{Fig.4}
\end{figure}

The coordinates  and spins of all minima are collected in Table I; these minima  are taken from the contour plots shown in Figs.2, 3 and 4, respectively.
\begin{table}[h!]
\begin{tabular}{|c|cc| c c c|c|}
\hline\\
quantization axis&$\theta_{min}$[rad]&$\varphi_{min}$[rad]&$I_1[\hbar]$&$I_2[\hbar]$&$I_3[\hbar]$&$H_{rot,min}$[MeV]\\
\hline
axis-1          &0.388 &0.8703&16.198& 4.269& 5.063&1.203\\
axis-2          &1.206 &1.236&15.443 &6.238& 5.370&1.381\\
axis-2          &1.104 &-0.983&- 13.003& 7.879& 8.666&2.960\\
axis-3&           1.124    &0.283  &15.152&4.403&7.569&  1.361       \\
\hline
\end{tabular}
\caption{Coordinates of the minima points for $H_{rot}$ and the corresponding values of the spin components.} 
\end{table}
Details about the behavior of $H_{rot}$ around its minima can be seen in Fig. 5, where the energy function is represented separately as a function of $\varphi$ and $\theta$, respectively, 
when the other variable is fixed in its minimum value.

It is worth mentioning that the stationary components of angular momentum given in the first row of Table I  describing the minimum of Fig. 1, coincide with those from the last row of 
Eq.(\ref{stavect}), which are obtained by solving the equations (2.9) and (2.7), and this happens despite the fact the two sets correspond to different spaces,one generated by the angular momentum components and one  is a two dimensional  phase space.

A major conclusion of this analysis is that irrespective of the chosen quantization axis, the deepest minimum of $H_{rot}$ is met for an angular momentum lying outside any principal plane of the inertia ellipsoid which is a prerequisite of a chiral motion.

The wobbling frequencies are determined by Eqs.(4.5), (4.9) and (4.13).Our calculations indicate that for the situation when the axis-1 is the quantization axis the system exhibits a minimum around which the system oscillates with a frequency equal to 0.245 MeV, while for axis-2 and axis-3 the system oscillates around the true minima with the frequency of 0.205 MeV and 0.065 MeV, respectively.

\begin{figure}[h!]
\includegraphics[width=0.3\textwidth,height=3cm]{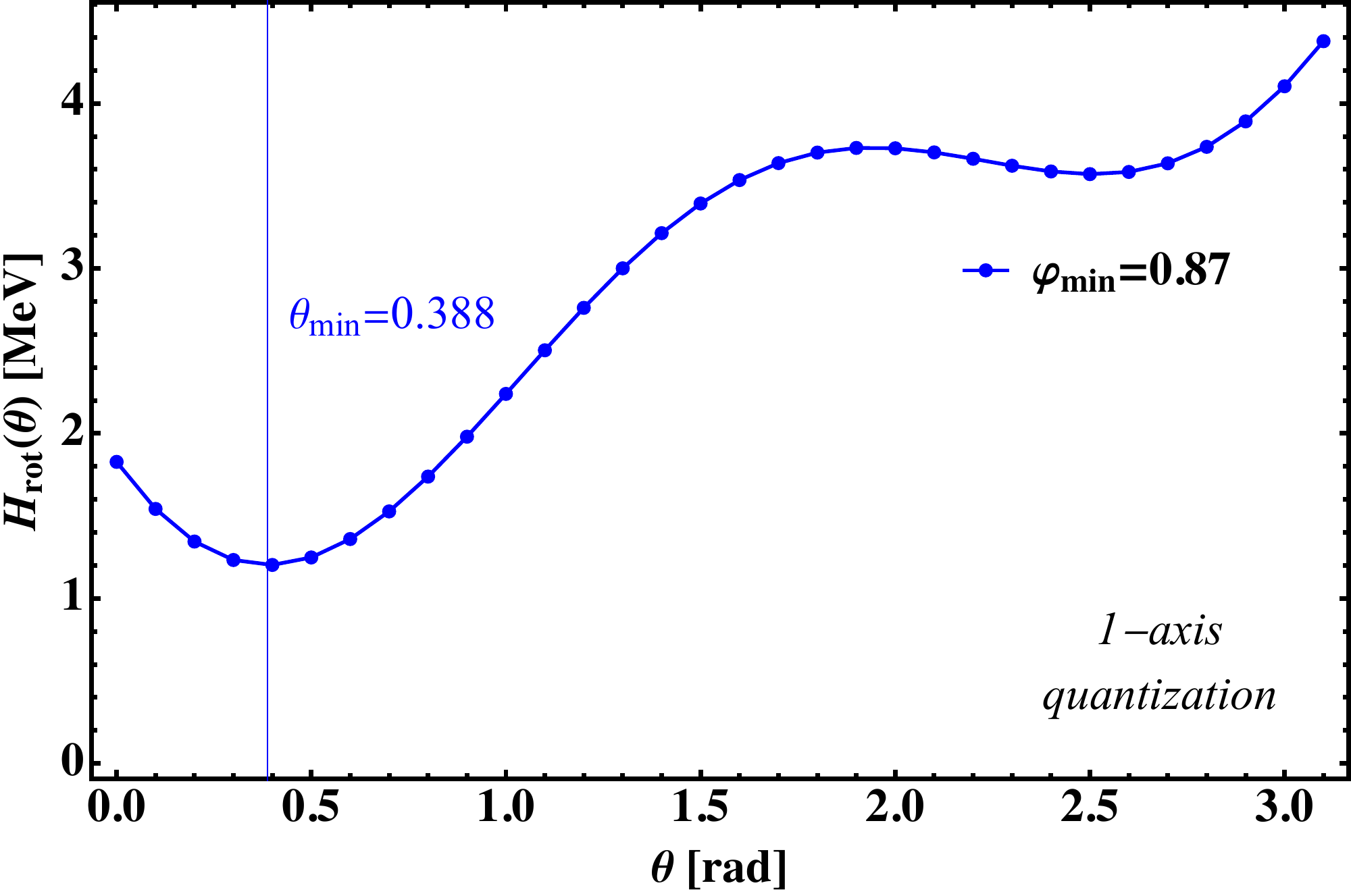}\includegraphics[width=0.3\textwidth,height=3cm]{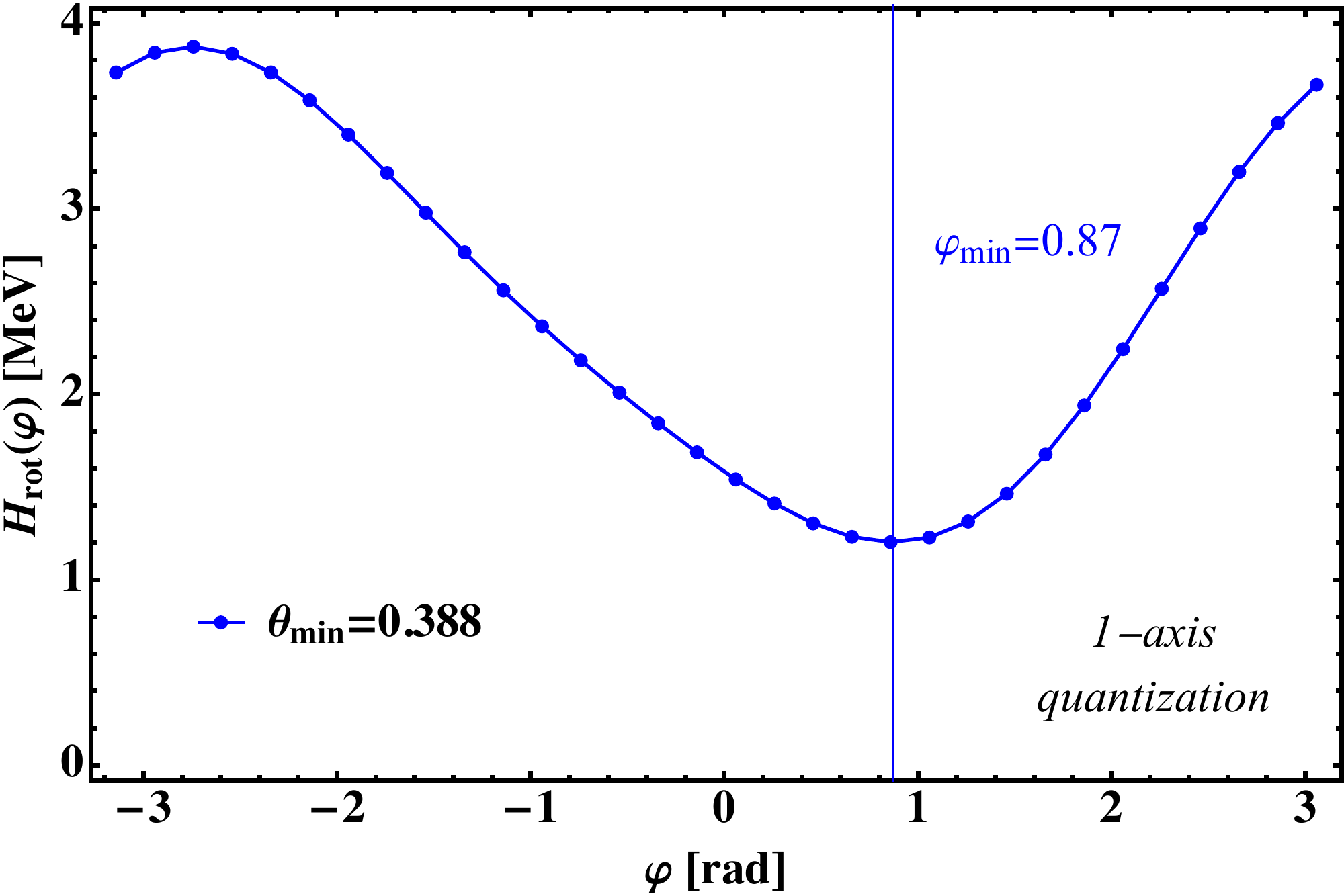}
\includegraphics[width=0.3\textwidth,height=3cm]{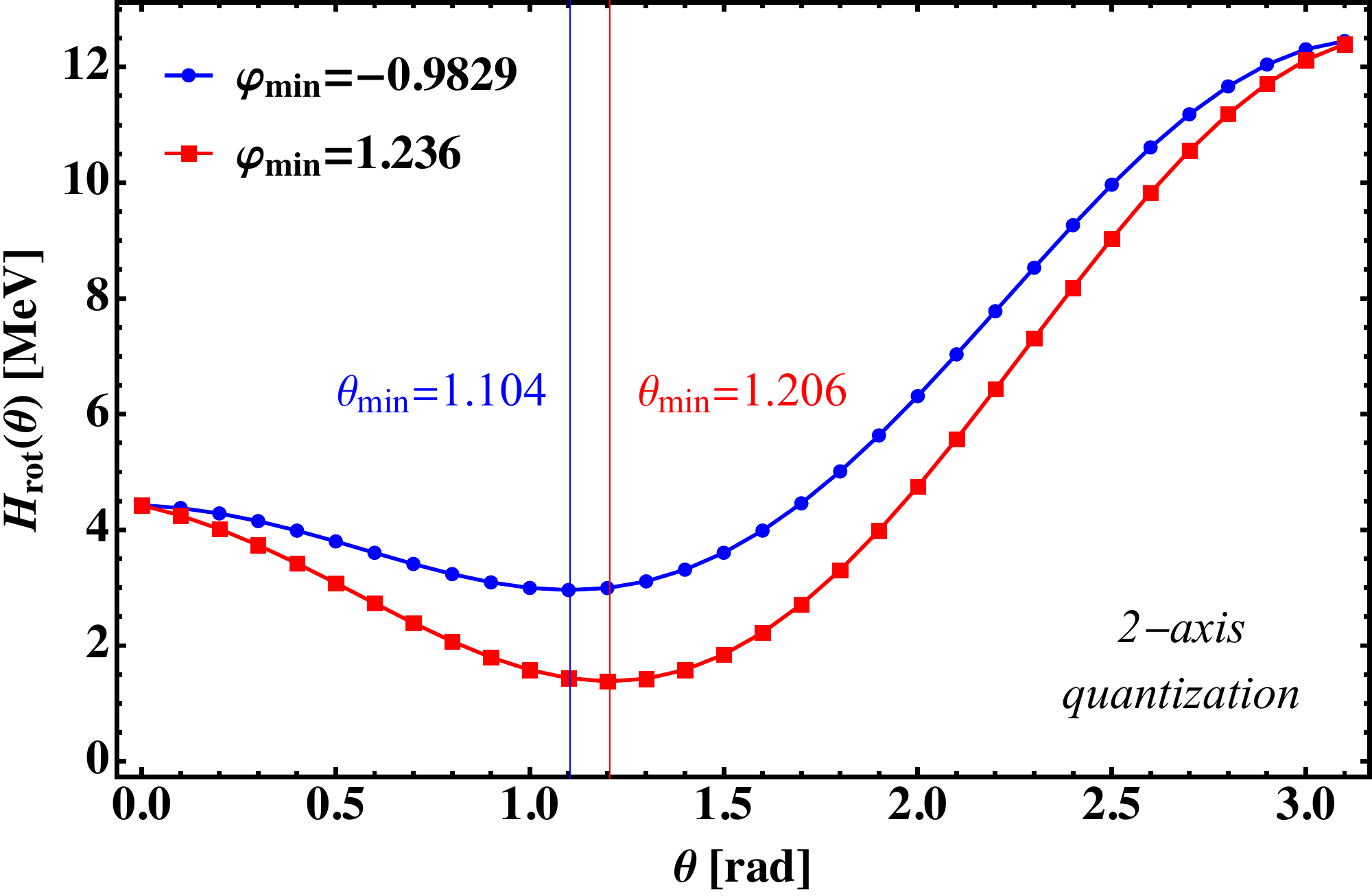}\includegraphics[width=0.3\textwidth,height=3cm]{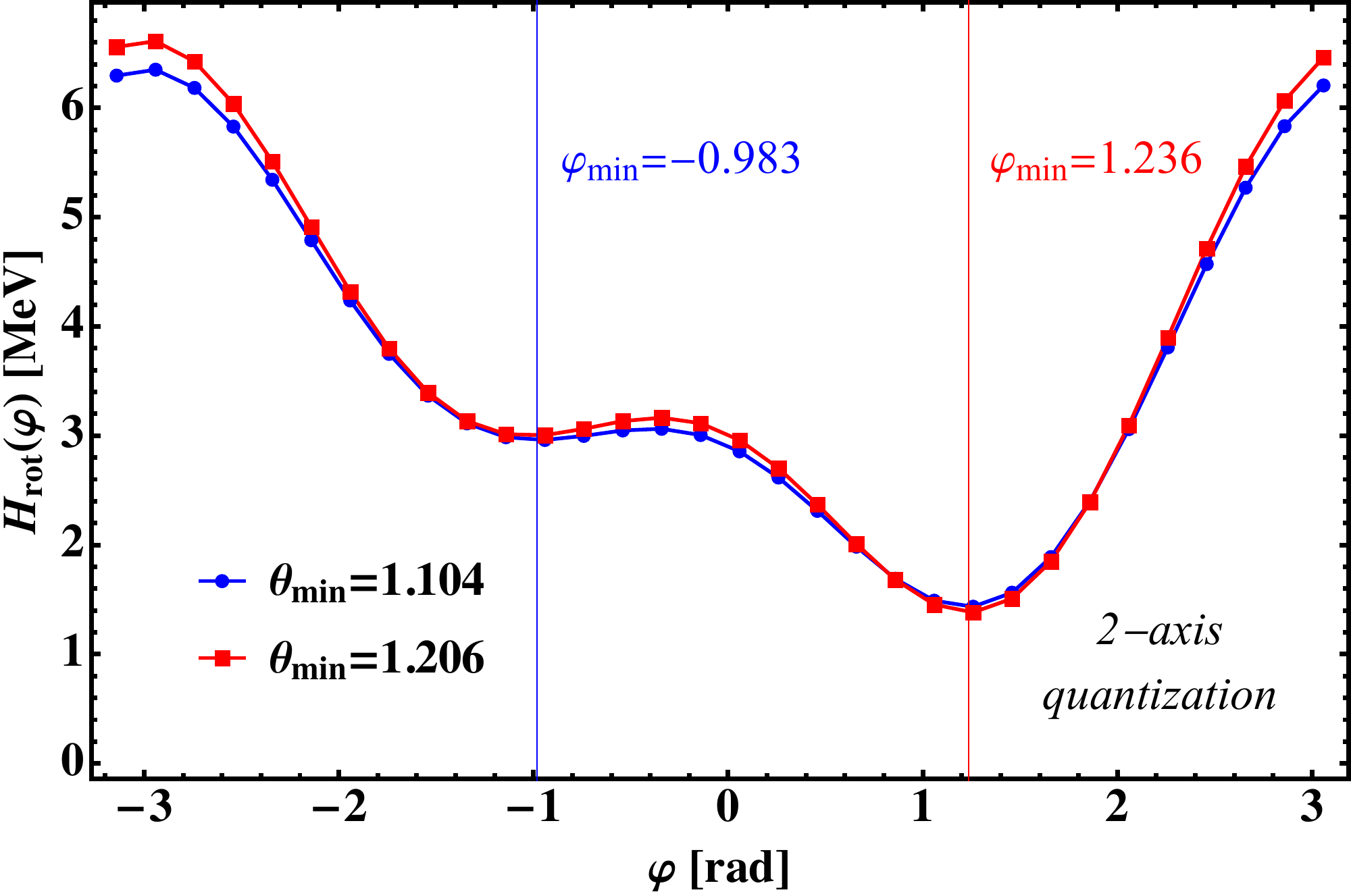}
\includegraphics[width=0.3\textwidth,height=3cm]{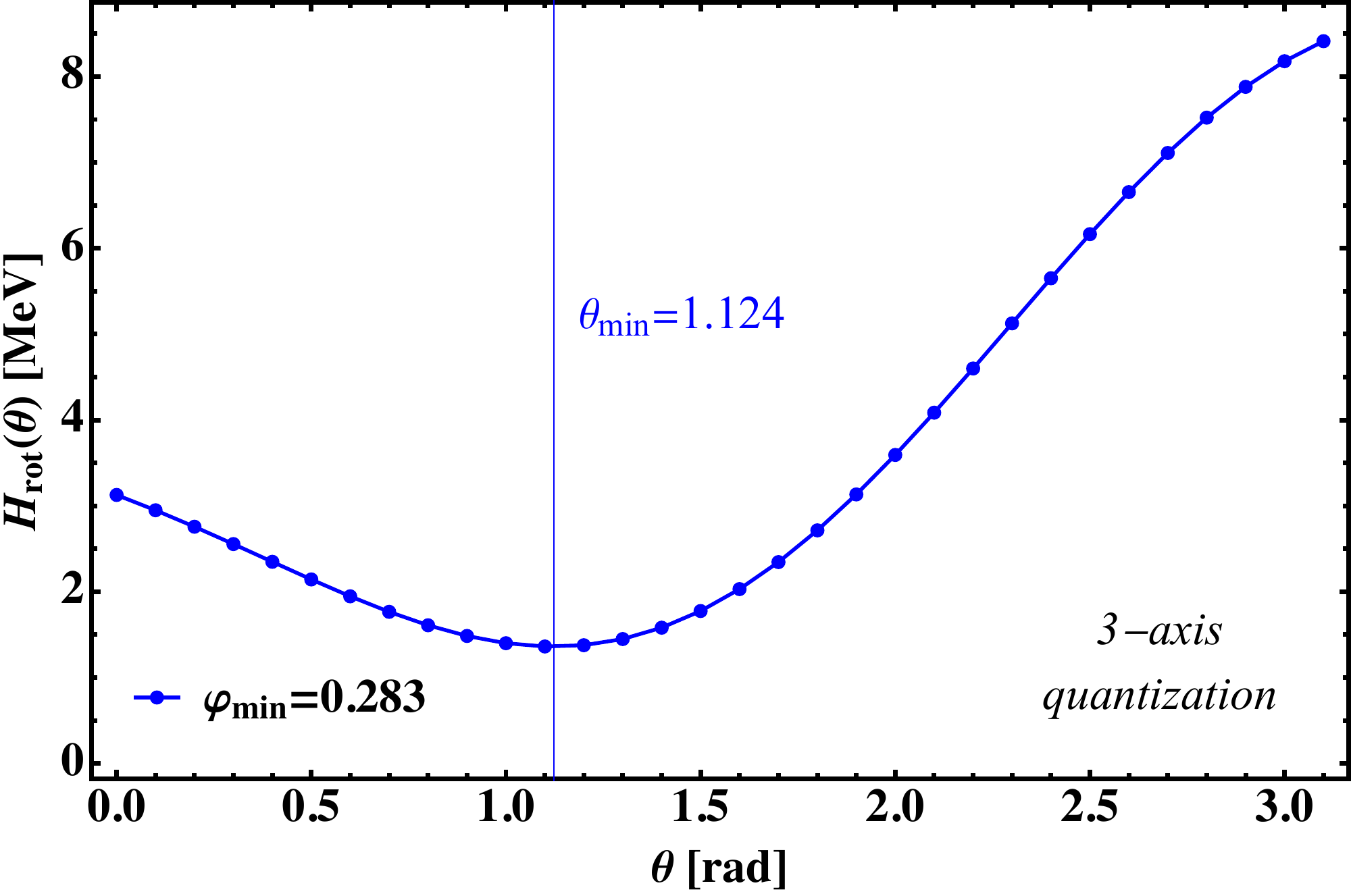}\includegraphics[width=0.3\textwidth,height=3cm]{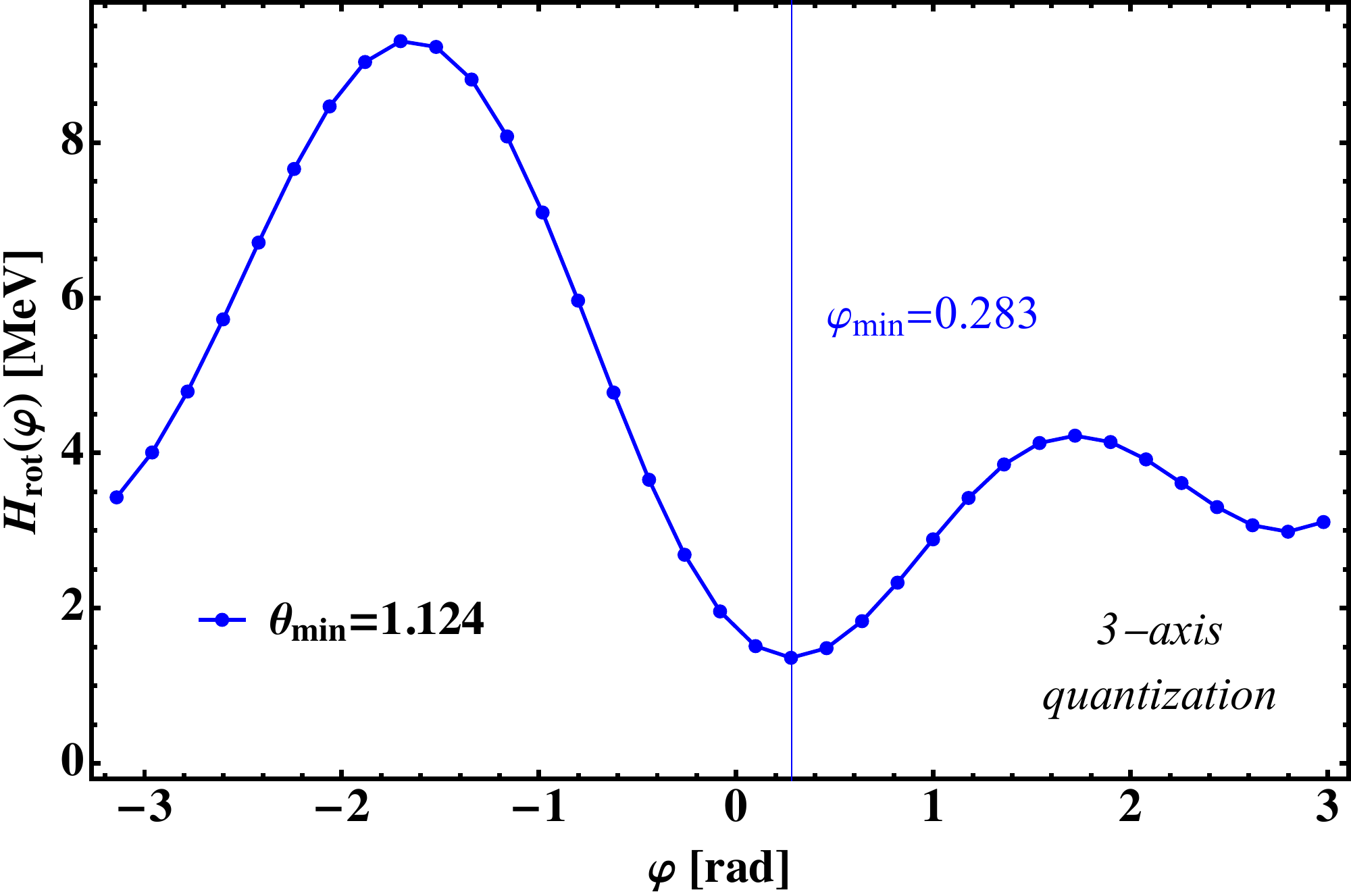}
\caption{(Color online).The classical energy function $H_{rot}$ as function of the angle $\varphi$ when $\theta$ is fixed at its minimum value as well as of function of $\theta$ for $\varphi$ taken in its minimal value. These plots are made for three distinct situations, namely when the quantization axis is axix-1 (first row), axis-2 (second row) and axis-3 (third row), respectively.}
\label{Fig.5}
\end{figure}

\begin{table}[h!]
\begin{tabular}{|c|cc| c c c|c|}
\hline\\
quantization axis&$\theta_{min}$&$\varphi_{min}$&$I_1[\hbar$&$I_2[\hbar]$&$I_3[\hbar]$&$H^{ch}_{rot,min}[MeV]$\\
\hline
axis-1          &2.753 &- 2.27 &- 16.198& - 4.269& - 5.063&1.202 \\
axis-1          &2.894 & 3.141 &- 16.965 &- 4.293 &$\approx$0.0&1.478 \\
axis-2          &1.935& - 1.905&- 15.443& - 6.238& - 5.370&1.381 \\
axis-2          & 2.148&3.141&$\approx$ 0& - 9.553 &- 14.662&2.873 \\
axis-3          &2.018& - 2.859& - 15.152& - 4.403& - 7.568&1.361 \\
axis-3          &2.021& 3.142 &- 15.754& $\approx$ 0& - 7.620&1.719 \\
\hline
\end{tabular}
\caption{Coordinates of the minima points for the chirally transformed Hamiltonian, $H^{ch}_{rot}$, and the corresponding values of the spin components} 
\end{table}
\begin{figure}[h!]
\begin{center}
\includegraphics[width=0.5\textwidth]{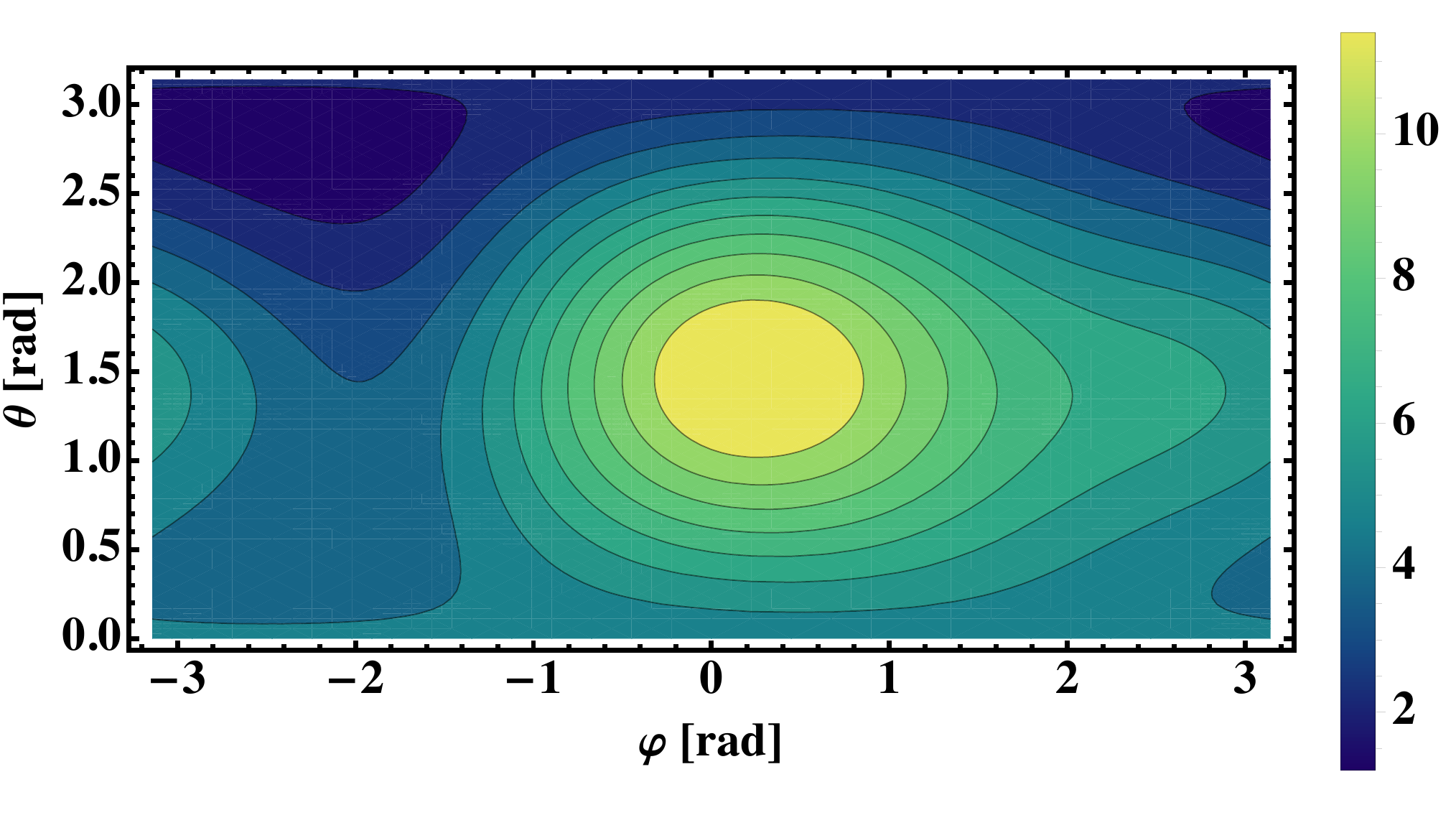}
\end{center}
\caption{(Color online) Contour plot when the axis-1 is the quantization axis. The energy function is $H^{ch}_{rot}$, the chiral image of the classical energy given by Eq. (2.2).}
\label{Fig.6}
\end{figure}
\begin{figure}[h!]
\begin{center}
\includegraphics[width=0.5\textwidth]{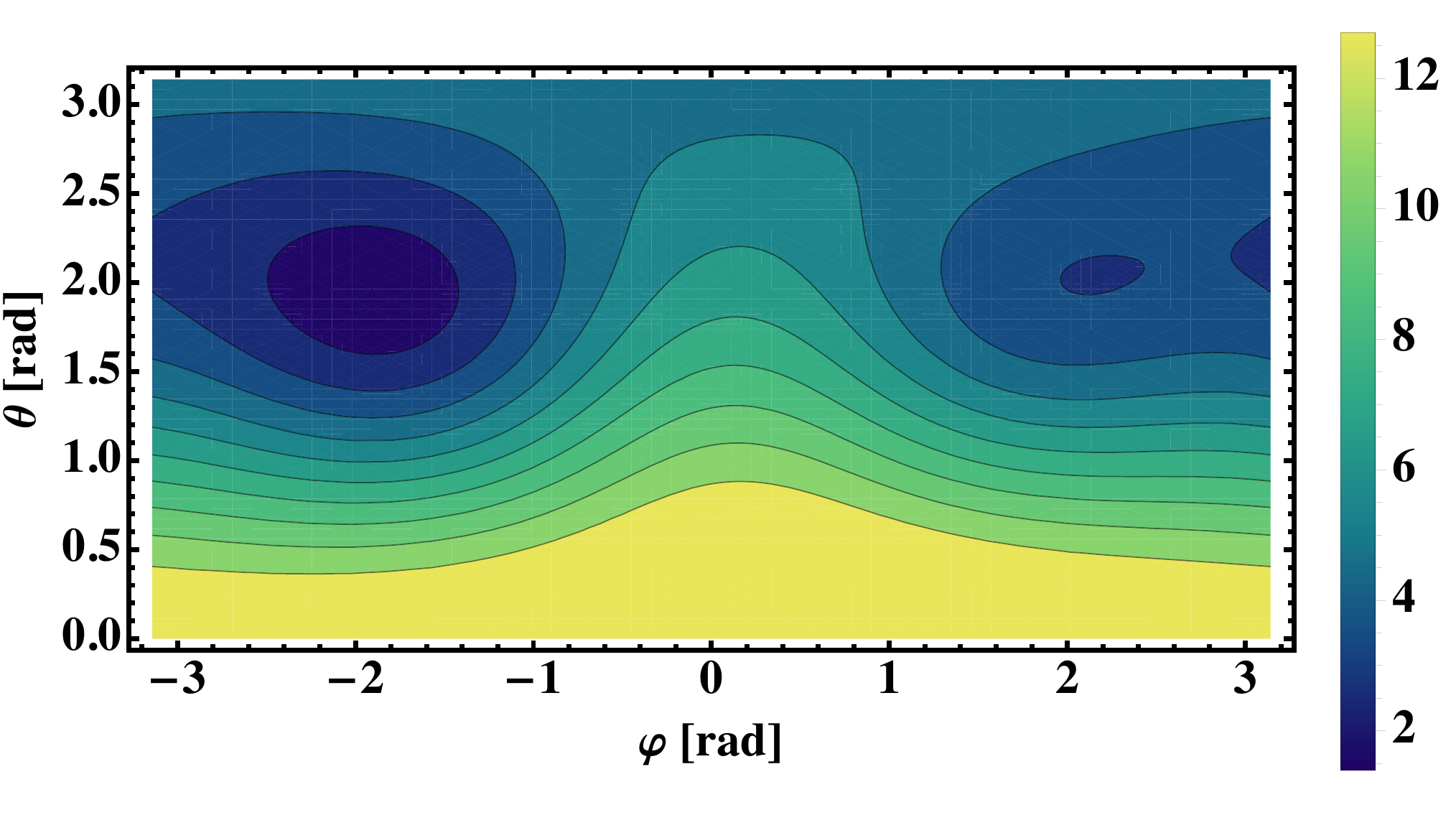}
\end{center}
\caption{(Color online) Contour plot when the axis-2 is the quantization axis. The energy function is  $H^{ch}_{rot}$, the chiral image of of the classical energy  given by Eq. (2.2).}
\label{Fig.7}
\end{figure}
\begin{figure}[h!]
\begin{center}
\includegraphics[width=0.5\textwidth]{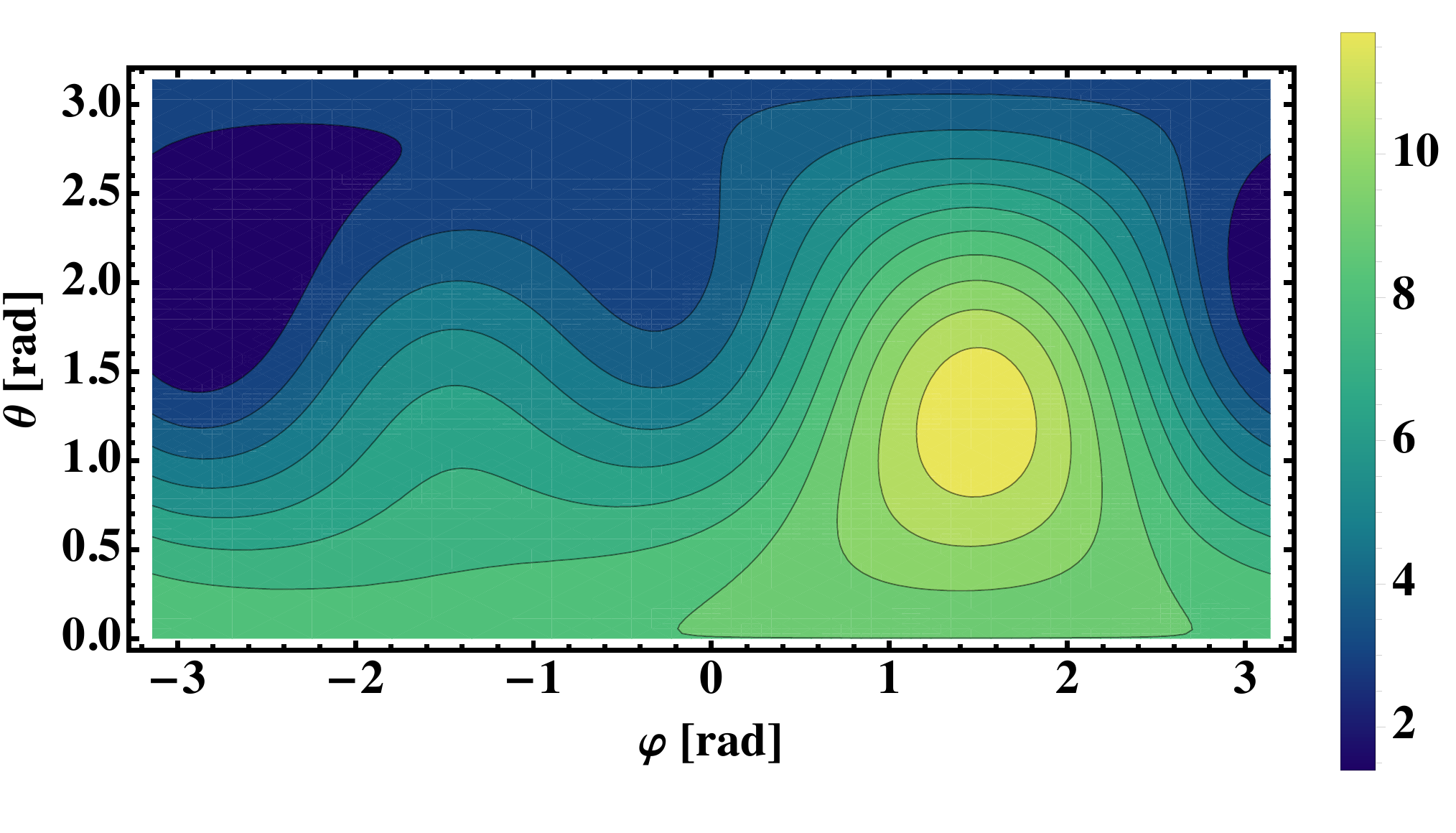}
\end{center}
\caption{(Color online) Contour plot when the axis-3 is the quantization axis. The energy function is  $H^{ch}_{rot}$, the chiral image of of the classical energy  given by Eq. (2.2).}
\label{Fig.8}
\end{figure}
The transformed Hamiltonian $H^{ch}_{rot}$ has the contour plots graphically represented in Figs. 6, 7 and 8 for the quantization axes 1, 2 and 3 respectively, with the minima coordinates collected in Table II.
Fixing one coordinate in its minimal value, $H^{ch}_{rot}$ becomes  a function of a single variable which exhibits several stationary points. These are plotted in Fig. 9. From there one can see that there are situations when  beside the main minimum the system exhibits several local minima.
\begin{figure}[h!]
\includegraphics[width=0.3\textwidth,height=3cm]{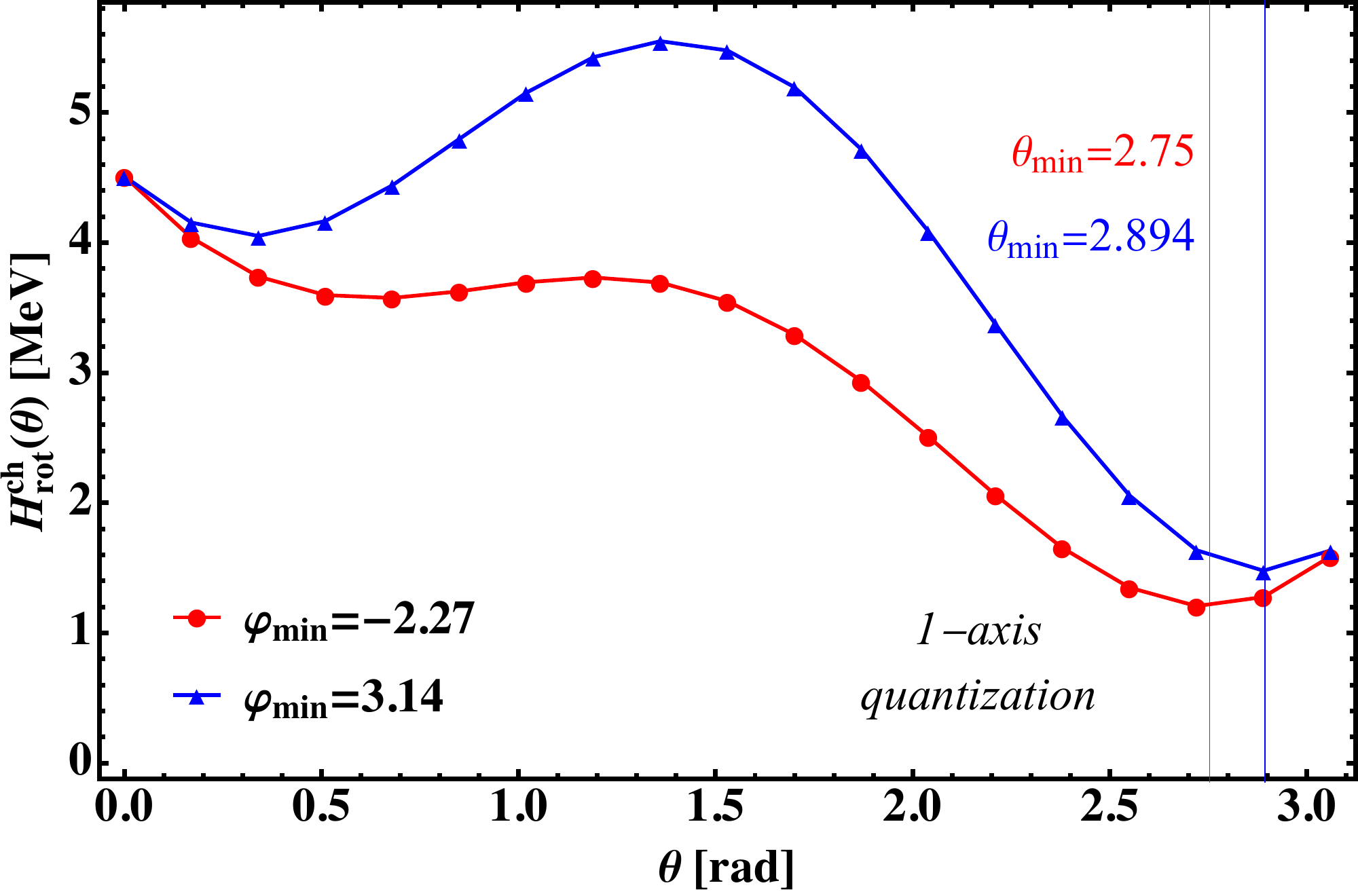}\includegraphics[width=0.3\textwidth,height=3cm]{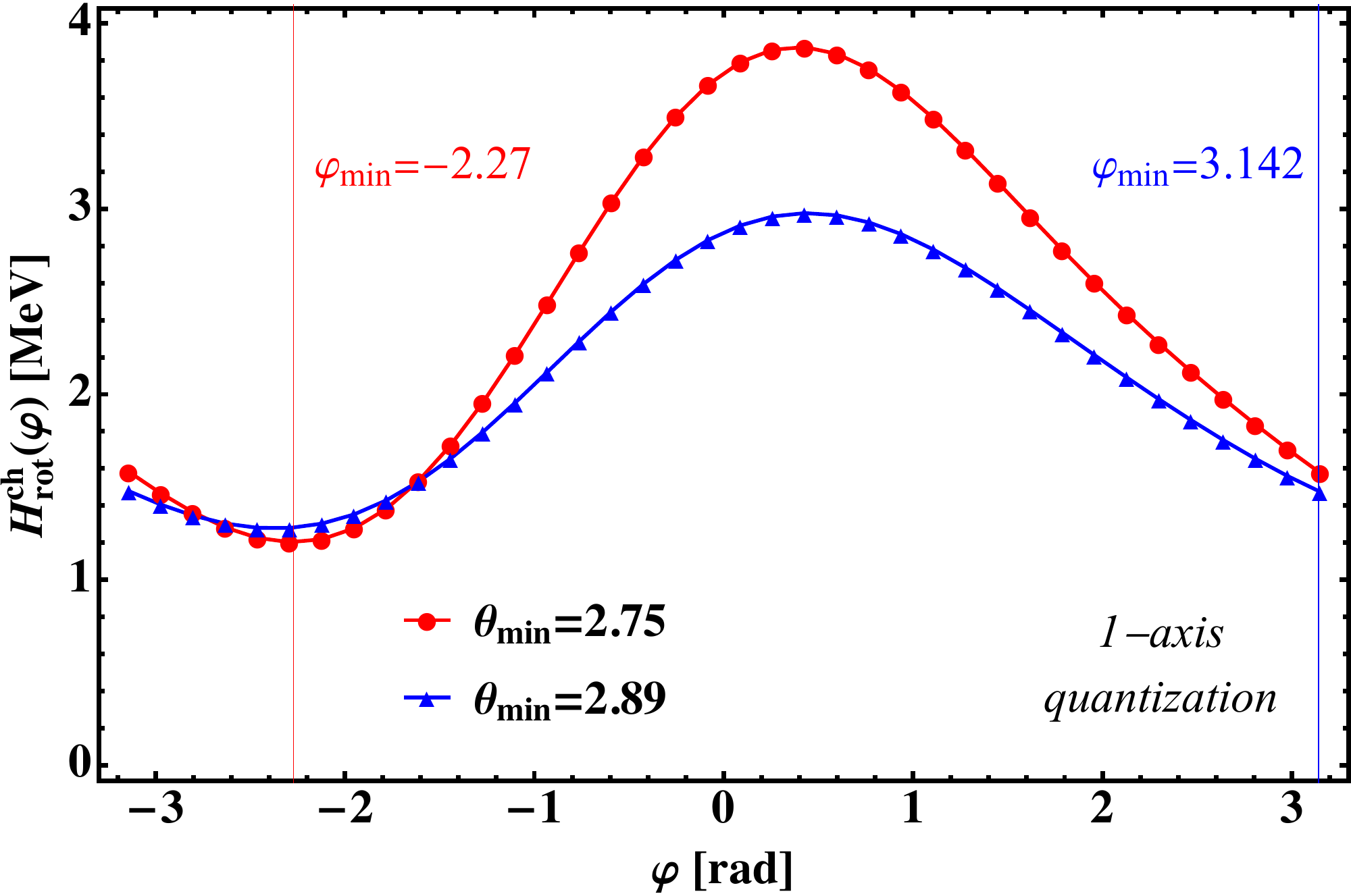}
\includegraphics[width=0.3\textwidth,height=3cm]{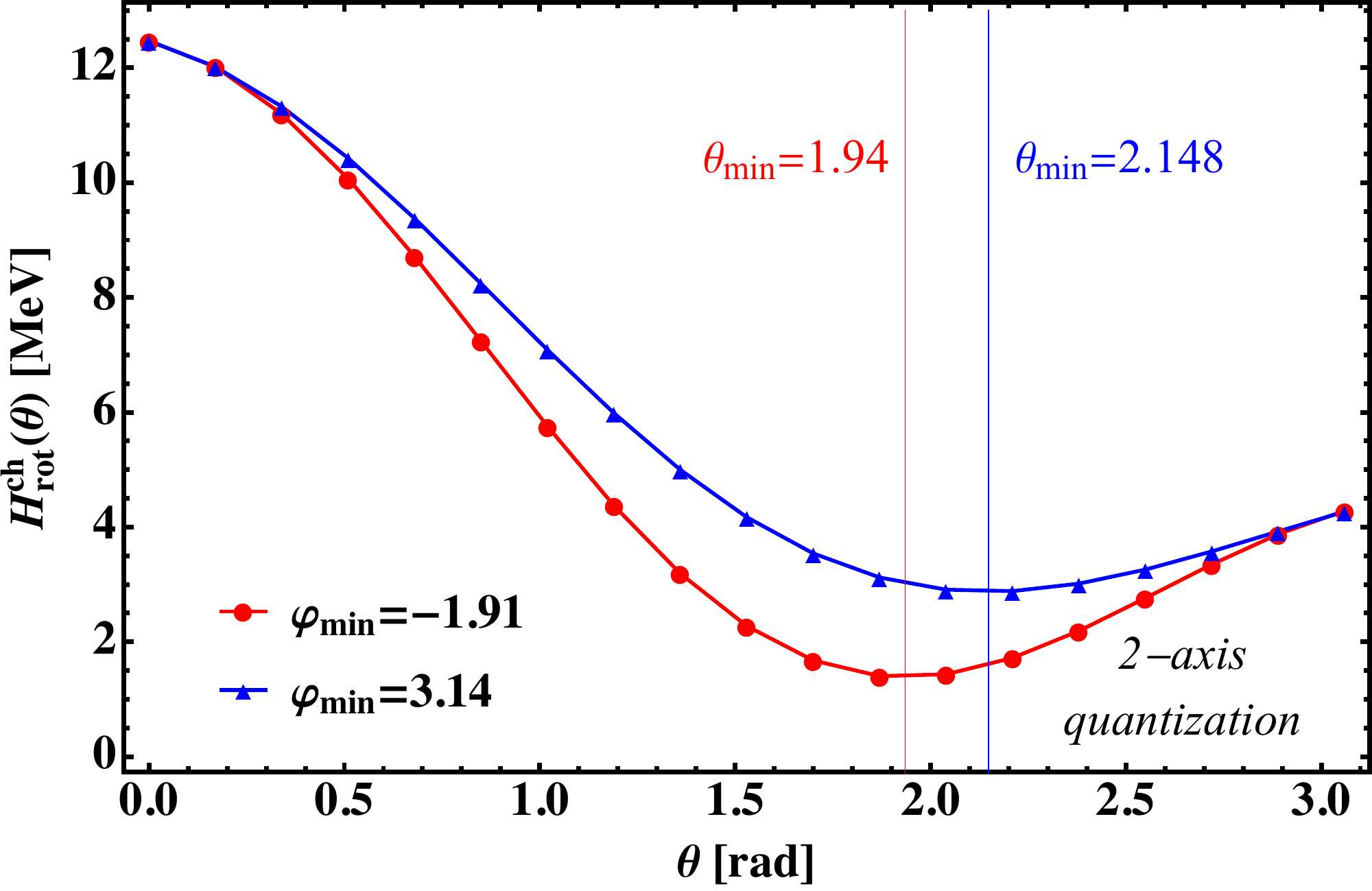}\includegraphics[width=0.3\textwidth,height=3cm]{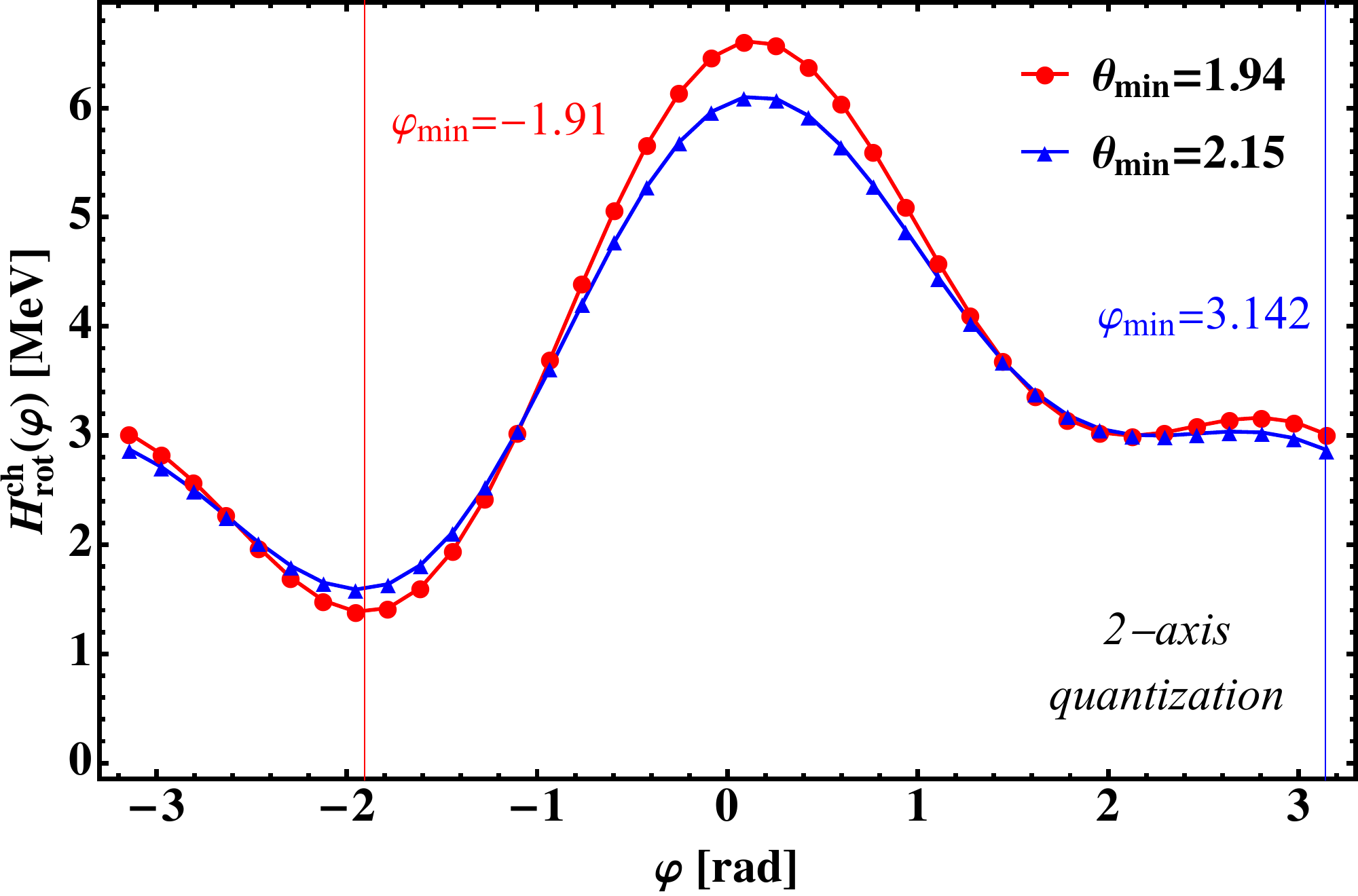}
\includegraphics[width=0.3\textwidth,height=3cm]{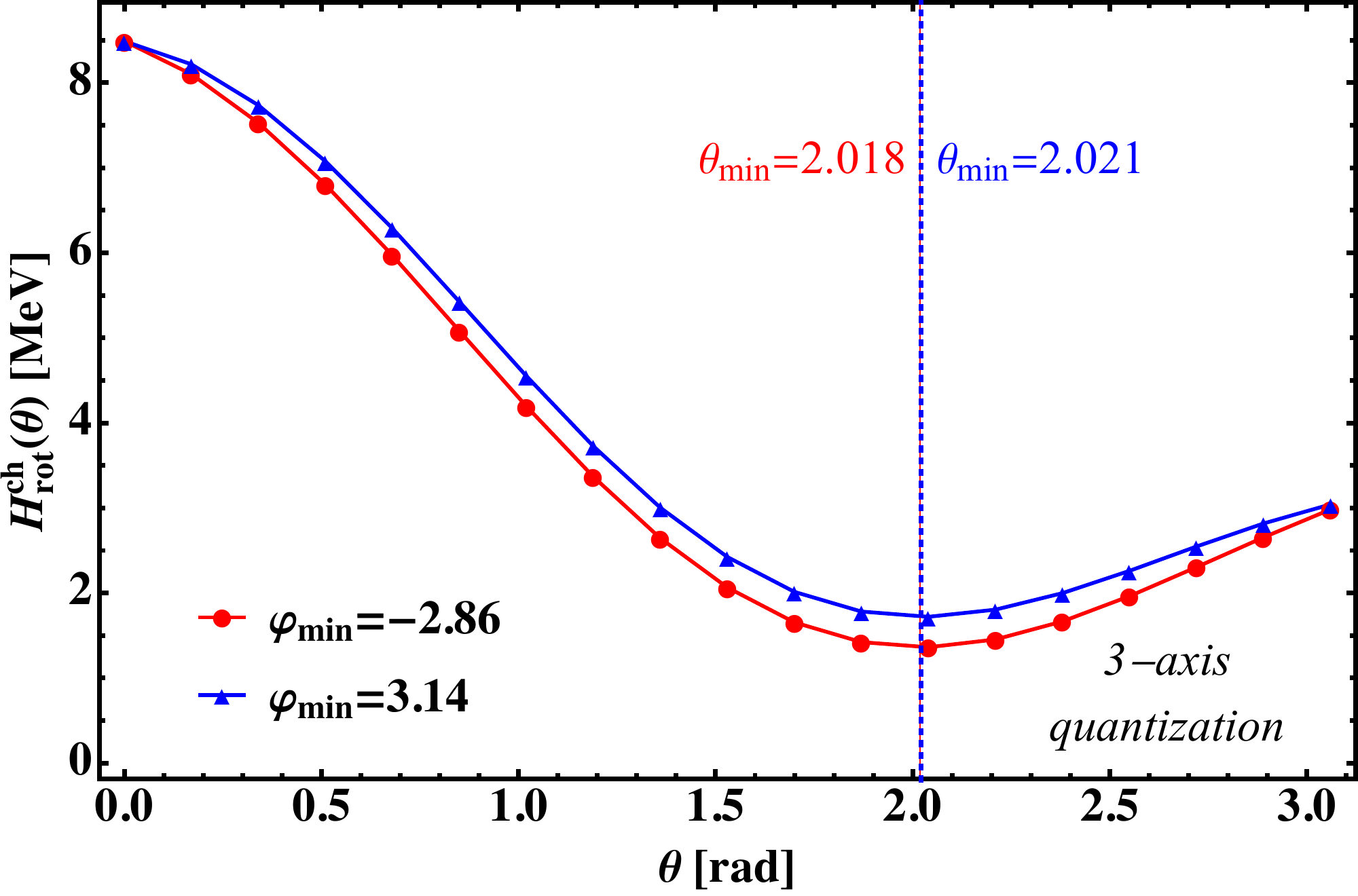}\includegraphics[width=0.3\textwidth,height=3cm]{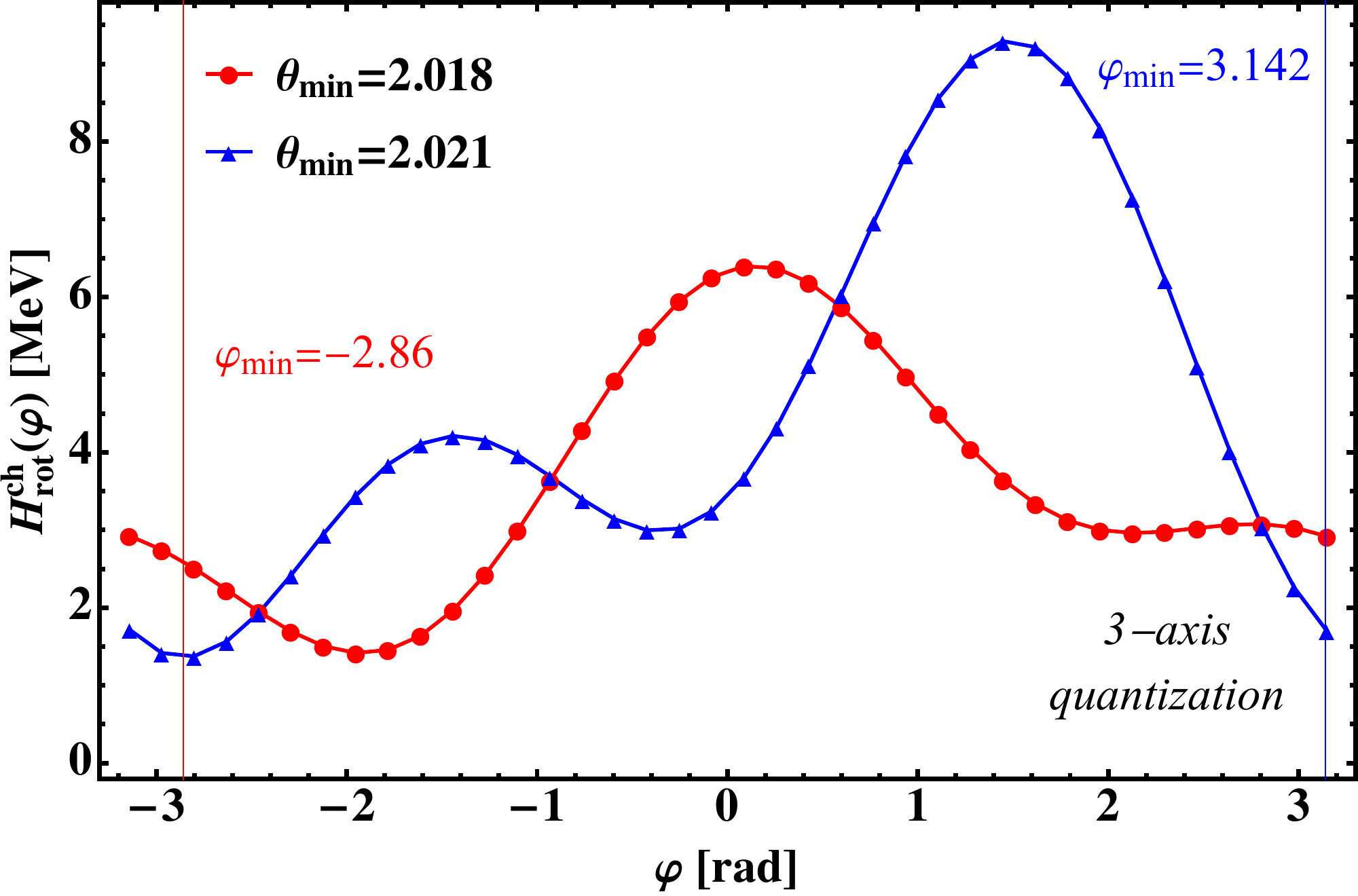}
\caption{(Color online).The classical energy function $H^{ch}_{rot}$ as function of the angle $\varphi$ when $\theta$ is fixed at its minimum value as well as of function of theta for $\varphi$ taken in its minimal value. These plots are made for three distinct situations, namely when the quantization axis is axix-1 (first row), axis-2 (second row) and axis-3 (third row), respectively.}
\label{Fig.9}
\end{figure}

The wobbling frequencies corresponding to the deepest minima when the quantization axis is the axis 1, 2 and 3 respectively, have the values:0.245, 0.209, 0.210 MeV, respectively.
One notes that when axis-1 is the quantization axis, the Hamiltonians $H_{rot}$ and $H^{ch}_{rot}$ have the same wobbling frequencies,  while for axis-2 as quantization axis the wobbling frequencies are very close to each other. Moreover the two Hamiltonians have the same/almost the same values in the respective minima. This is a reflection of the fact that for the two cases the chiral invariant part of $H_{rot}$ prevails over the non-invariant part.
Consequently, for the cases when the quantization axis is the axis-1 or 2, one can build up two wobbling bands with a similar structure which results in having a twin pair of bands.

 Note that contrary to the previous publications, where the odd nucleon was  rigidly fixed either to an axis \cite{Buda} or to a principal plane \cite{Rad2020} of the inertia ellipsoid, here the rigid coupling is achieved to a direction which does not belong to a principal plane. In this way one conciliates between the two signatures of triaxial nuclei,  these being simultaneously considered.

\newpage  
\setcounter{equation}{0}
\renewcommand{\theequation}{7.\arabic{equation}}
\section{Conclusions}
In the previous section we developed a classical interpretation of the wobbling motion of an even-odd system described by a particle-triaxial rigid core coupling. The odd particle is rigidly coupled to the deformation and the total angular momentum such that its angular momentum lays outside any principal plane of the inertia ellipsoid.
Equations of motion for the angular momentum components are studied both in the space of angular momenta and in the reduced space of the generalized phase space coordinates.
By a quadratic expansion of the classical energy function around a stationary point one finds the analytical expression of the wobbling frequency. The same procedure was applied also for the chirally transformed Hamiltonian. Formalism was applied to an illustrative example. One found out that there exist  minimum points for the energy function where the model Hamiltonian and its chiral image admit real wobbling frequencies.Extending the calculations to a set of total angular momenta one certainly obtains a pair of chiral twin doublet band with similar properties.
Concluding, this work provides an inedited picture of triaxial nuclei, by being able to potentially unify two signatures in a consistent manner. Moreover the semi-classical treatment of the problem is indeed a remarking for the proposed model. To our knowledge, there are no other approaches within the literature that aim at describing both phenomena simultaneously. Nevertheless, new experimental data is necessary for testing grounds. 

{\bf Acknowledgment.} This work was supported by the Romanian Ministry of Research and Innovation through the project PCE nr. 16/2021

\setcounter{equation}{0}
\renewcommand{\theequation}{A.\arabic{equation}}
\section{Appendix A}
The solutions for Eq.(\ref{eqomeg}) are given by the Cardano formulas:
\begin{eqnarray}
\omega_1&=&s_1+s_2,\nonumber\\
\omega_2&=&-\frac{1}{2}(s_1+s_2)+\frac{i\sqrt{3}}{2}(s_1-s_2),\nonumber\\
\omega_3&=&-\frac{1}{2}(s_1+s_2)-\frac{i\sqrt{3}}{2}(s_1-s_2).
\end{eqnarray}
where the following notations have been used:
\begin{eqnarray}
s_1&=&\left(T+(T^2+S^3)^{1/2}\right)^{1/3},\nonumber\\
s_2&=&\left(T-(T^2+S^3)^{1/2}\right)^{1/3}.
\end{eqnarray}
If $T^2+S^3>0$,at least one solution is real. If $T^2+S^3<0$ all solutions are imaginary, while if $T^2+S^3=0$ solutions are real and at least two are equal.


\begin{thebibliography}{99}
\bibitem{Davy}A.S. Davydov and G.F. Filippov, Nucl. Phys. {\bf 8} (1958) 788.
\bibitem{Terven}J. Meyer-ter-Vehn, F. S. Stephens and R.M. Diamond, Phys. Rev. Lett. {\bf 32} (1974)1383;
J. Meyer-ter-Vehn, Nucl. Phys. {\bf A249} (1975) 111, 141.
\bibitem{WilJean} L. Wilets and M. Jean, Phys. Rev. {\bf 102}, (1956)788.
\bibitem{Marsh}E. R. Marshalek, Nucl. Phys. {\bf A 331} (1979) 429.
\bibitem{Fra} St. Frauendorf, Rev. Mod. Phys., vol. {\bf 73},  (2001)463.
\bibitem{Bona} D. Bonatsos {\it et al.} Phys. Lett. {\bf B 584} (2004) 40.
\bibitem{Bona2} D. Bonatsos, D. Lenis, D. Petrellis and P. A. Terziev, Phys. Lett. {\bf B 588},(2004)172.
\bibitem{Bona3} D. Bonatsos, D. Lenis, D. Petrellis, P. A. Terziev and I. Yigitoglu, Phys. Lett. {\bf B 621} (2005) 102.
\bibitem{BoMo}A. Bohr and B. Mottelson, {\it Nuclear Structure} (Benjamin, Reading, MA, 1975), Vol. II, Ch. 4.
\bibitem{Rad016}A. A. Raduta, Progress in Particle and Nuclear Physics 90 (2016) 241.
\bibitem{Petrache96}C. M. Petrache {\it et al.,} Nucl. Phys. {\bf A597} (1996) 106.
\bibitem{Petrache06}C. M. Petrache, {\it et al.}, Phys. Rev. Lett. {\bf 96} (2006) 112502.
\bibitem{Frau97}S. Frauendorf and J. Meng, Nucl. Phys. {\bf A 617} (1997) 131.
\bibitem{Frau016}S. Frauendorf and F. Donau, Phys. Rev. C{\bf 89}(2014) 014322.
\bibitem{Ham} I. Hamamoto, Phys. Rev. {\bf C 65} (2002) 044305.
\bibitem{Ham1}I. Hamamoto and G. B. Hagemann, Phys. Rev. {\bf C 67} (2003) 014319.
\bibitem{Tan017} K. Tanabe and K. Sugawara Tanabe, Phys. Rev. C {\bf 95} (2017) 064315.
\bibitem{Frau018} S. Frauendorf, Phys. Rev. {\bf C97}(2018) 069801.
\bibitem{Tana018} K. Tanabe and K. Sugawara Tanabe, Phys. Rev. C {\bf 97},(2018) 069802 .
\bibitem{Rad07} A. A. Raduta, R. Budaca and C. M. Raduta, Phys. Rev. {\bf C 76}, (2007) 064309.
\bibitem{Rad017} A. A. Raduta, R. Poenaru and L. Gr. Ixaru, Phys. Rev. C {\bf 96},(2017) 054320.
\bibitem{Rad018} A. A. Raduta, R. Poenaru and Al. H. Raduta, J. Phys. G: Nucl. Part. Phys. {\bf 45},(2018) 105104 .
\bibitem{Rad20} A. A. Raduta, R Poenaru and C M Raduta, J. Phys. G: Nucl. Part. Phys. {\bf 47},(2020) 025101 .
\bibitem{Rad201} A. A. Raduta, R Poenaru and C M Raduta, Phys. Rev. C {\bf 101},(2020)014302 .
\bibitem{Rad2020} A.A.Raduta, C.M. Raduta, and R. Poenaru,J. Phys. G: Nucl. Part. Phys. {\bf 48}, (2020) 015106.
\bibitem{Buda} R. Budaca, Phys. Rev. C {\bf 97} (2018) 024302.
\bibitem{David} A.C. Davydov, {\it Teoria atomnova yadra}, Moscva, 1958 (in russian), chapters 19,20.
\bibitem{Ring} P. Ring and P. Schuck,{\it The Nuclear Many Body Problem} Springer-Verlag, Berlin, Heidelberg 2000, p.110
\bibitem{Beng}H. Frisk and  R. Bengston, Phys. Lett. {\bf 196 B} (1987) 14.
\end{thebibliography}
\end{document}